\documentclass[twocolumn]{aastex63}
\usepackage{color}
\usepackage{xcolor}
\usepackage[normalem]{ulem}
\usepackage[caption=false]{subfig}

\newcommand{\srcname}{RGB1}
\newcommand{\lya}{Ly$\alpha$}

\received{September 7, 2023}
\revised{May 2, 2024}
\accepted{May 23, 2024}

\shorttitle{Radio Emission in a Green Bean Galaxy}
\shortauthors{Sanderson et al.}

\graphicspath{{./}{figures/}}

\begin{document}

\title{Signatures of AGN feedback modes: A Green Bean Galaxy with\\ 150~kpc jet-induced radio emission}

\author[0000-0001-5825-7683]{Kelly N. Sanderson}
\altaffiliation{Grote Reber Fellow; ksand017@nmsu.edu}
\affiliation{Department of Astronomy, New Mexico State University, P.O.Box 30001, MSC 4500, Las Cruces, NM, 88033, USA}
\affiliation{National Radio Astronomy Observatory, Domenici Science Operations Center, PO Box 0, Socorro, NM 87801, USA}

\author[0000-0002-5289-5729]{Anna D. Kapi{\'n}ska}
\affiliation{National Radio Astronomy Observatory, Domenici Science Operations Center, PO Box 0, Socorro, NM 87801, USA}

\author[0000-0001-8302-0565]{Moire K. M. Prescott}
\affiliation{Department of Astronomy, New Mexico State University, P.O.Box 30001, MSC 4500, Las Cruces, NM, 88033, USA}

\author[0009-0003-6389-4101]{Audrey F. Dijeau}
\affiliation{Department of Astronomy, New Mexico State University, P.O.Box 30001, MSC 4500, Las Cruces, NM, 88033, USA}

\author[0000-0002-1313-429X]{Savannah R. Gramze}
\affiliation{Department of Astronomy, University of Florida, Gainesville, FL 32611 USA}

\author[0009-0002-2501-6706]{Jacqueline Hernandez}
\affiliation{Space Telescope Science Institute, Baltimore, MD 21218, USA}

\author[0000-0001-7511-4836]{Katherine T. Kauma}
\affiliation{Institute of Astronomy, University of Cambridge, Madingley Road, Cambridge CB3 0HA, UK }

\begin{abstract}
Jetted Active Galactic Nuclei (AGN)  hosting extended  photoionized nebulae provide us with a unique view of the timescales associated with AGN activity. Here, we present a new Green Bean galaxy at $z=0.304458\pm0.000007$ with large scale  jet-induced radio emission. The Spectral Energy Distributions (SEDs) of the radio components show steep spectral indices ($\alpha=-0.85$ to $-0.92$ for the  extended regions, and $\alpha=-1.02$ for the faint radio core), and spectral age modeling of the extended radio emission indicates that the lobes are  $>$6~Myrs old.  It is unclear whether the jet is active, or remnant with an off-time of 2-3~Myr. Several detached clouds lie around the host galaxy up to 37.8~kpc away from the nucleus, and their ionization profile indicates a decline ($\sim$2~dex) in the AGN ionizing photon production over the past $\sim$0.15~Myr. Furthermore, we  measure a blue shift for one of the clouds  that is spatially coincident with the path of the radio jet. The cloud  is likely illuminated by the photoionizing AGN, and potentially underwent an interaction with the relativistic jet. Our multiwavelength analysis suggests that \srcname\ was in a phase of jet production prior to the radiatively efficient accretion phase traced by the detached cloud emission. It is unclear whether \srcname\ transitioned into a low-excitation radio galaxy or an inactive galaxy over the past $\sim$0.15~Myr, or whether the extended radio and optical emission trace distinct accretion phases that occurred in succession.

\end{abstract}


\section{Introduction} 
The circumgalactic medium (CGM) forms the boundary between a galaxy and the intergalactic medium (IGM). Super-massive black hole (SMBH) accretion and active galactic nuclei (AGN) feedback play a role in galaxy evolution, and may leave imprint on the CGM gas as well as environments around the galaxies. Two main types (modes) of AGN feedback are recognized. The radiative mode feedback caused by radiation-driven winds is typical of the most luminous AGN at high redshift, while the kinetic mode feedback caused by radio jets transporting energy and material into the host galaxy surroundings is typical of low redshift jetted AGN  \citep{Silk1998, McNamara2012}. Radiative mode feedback is thought to accompany episodes of radiatively efficient high-Eddington accretion, while kinetic mode feedback is associated with episodes of radiatively inefficient low-Eddington accretion, and the production of radio jets and associated structures. The duration of the timescales associated with these modes of feedback are an important factor in determining how the gas supply available to galaxies is regulated. 

Information about the timescales associated with the duty cycle of AGN can be easily extracted from studies of Type 2 AGN hosting extended gaseous nebulae because our view of the host galaxy and its ``extended emission line region'' \citep[or EELR;][]{Stockton1983} is not hidden by the bright and energetic accretion disk. Light-travel times can provide lower limits on the duration of AGN activity by telling us that the AGN must have been producing ionizing photons for a time that is at least as long as the light travel time between the central AGN and the maximum extent of the EELR. If the production of ionizing photons in the central engine has declined more than 0.001-0.01 Myrs ago, the sources appear as an ionization echo where the extended gas around the AGN shows signatures of AGN activity but the nucleus of the host galaxy lacks evidence of active accretion \citep{Keel2012a, Oppenheimer2013, Esparza-Arredondo2020}. In these systems, we also obtain constraints on how long ago the central source of ionizing photons switched off \citep{Lintott2009, Schawinski2010, Keel2012a}.

Additional information about the timescales associated with AGN duty cycles is revealed when the Type 2 AGN simultaneously host radio jets. The population of accelerated charged particles that have been transported by the jets into the circumgalactic and intergalactic media will begin to radiate away their energy, with the higher energy charged particles radiating their energy away faster than the lower energy ones. This aging of the synchrotron-emitting radio plasma manifests as a curvature in the broadband radio spectrum of the radio structures. Modeling this aging plasma using spectral aging analysis \citep{Kardashev1962, Jaffe1973, Tribble1993, Harwood2013} allows us to infer the time since the plasma was last accelerated by the magnetic field.  In the case of active radio jets, this provides information on the duration of the jet production phase. For `jetted' AGN that have slowed their accretion rate and entered a dormant phase of radio activity, this type of modeling can also help constrain the time elapsed since the end of the jet production phase. By using light-travel times, and in the case of jetted Type 2 AGN spectral aging analyses, we can infer the timescales over which AGN activity varies in these sources. 

Comprehensive knowledge of the AGN activity requires understanding the relationship between the duty cycle and the host galaxy properties. One interesting sample that shows evidence for AGN that have undergone recent changes in their accretion state (ramped down) was found by \citet[][]{Schirmer2013}. ``Green Bean" (GB) galaxies are a class of nebulous objects,  discovered within the Sloan Digital Sky Survey (SDSS) DR8 out to z$\sim$0.6, that have strong spatially extended [OIII]$\lambda$5007 emission which has been redshifted into the r-band filter, causing them to appear elongated and bright green in optical imaging \citep{Schirmer2013}. These objects have been shown to emit enough \lya\ emission to classify them as low-z \lya\ nebulae \citep{Schirmer2016} and have bluer continua compared to other Type 2 AGN \citep{Prescott2019}, consistent with recent central star formation. So far only 17 of these objects have been identified in all of SDSS, making them a rare occurrence ($4.4/{\rm Gpc}^3$), and supporting the idea that these sources may be experiencing a short-lived phase in their evolution.

\citet{Kapinska2017} serendipitously discovered a GB-like object with extended radio emission in a search for radio galaxies with different Fanaroff-Riley \citep[FR;][]{Fanaroff&Riley1974} classifications on opposite sides of the nucleus \citep[``HYbrid MOrphology Radio Source'' or HyMoRS;][]{Gopal-Krishna2000}, using the Radio Galaxy Zoo \citep[RGZ;][]{Banfield2015}. The source, RGZ~J123300.2+060325 (hereafter \srcname) was found to meet 11 out of 12 of the \cite{Schirmer2013} color cuts used to identify the mostly non-jetted sample of GB galaxies, show no clear radio core in shallow FIRST imaging (potentially consistent with being ramped-down), and have extended radio morphology with a full projected linear size of $\sim 250\pm 30$kpc \citep{Kapinska2017}. Although this source was not included in the original \cite{Schirmer2013} GB sample (due to missing one colorcut) the source does pass the updated GB galaxy color selection criteria of \citet{Schirmer2016}. 

In this paper, we use low-frequency, high resolution follow-up radio observations from the Karl G. Jansky Very Large Array (VLA) to confirm the association of the radio galaxy core with the spatially coincident optical host galaxy, constrain the spectral age of the extended radio components, and investigate its magnetic field in an attempt of disentangling its radio morphology. Furthermore, we obtain new medium-resolution optical spectra using the Apache Point Observatory (APO) 3.5m Dual Imaging Spectrograph (DIS) spectrograph to investigate the kinematics of the host galaxy and its EELR, and investigate the morphology of the [OIII]-emitting gas along the radio axis of the source. We also obtain new low-resolution optical spectra from APO's Kitt Peak Ohio State Multi-Object Spectrograph (KOSMOS\footnote{KOSMOS was converted for use at APO in 2021}) to constrain the source redshift and the mechanisms powering the EELR. Through these analyses we attempt to reconstruct the evolutionary path of the galaxy over the past $\sim$10 Myr.

The paper is structured as follows. In Section \ref{sec:method}, we present our observations, and data reduction. The results are presented in Section \ref{sec:results}. In Section \ref{sec:discussion}, we discuss these results and their implications for the accretion history of the AGN in RGB1. We assume the standard $\Lambda$CDM cosmology, i.e., $\Omega_m=0.3$, $\Omega_{\Lambda}=0.7$, $h=0.7$. The angular scale at $z=0.305$ is 4.51~kpc/$^{\prime\prime}$.

\section{Observations \& Data Reduction} \label{sec:method}

\subsection{Radio Observations}
The VLA radio observations were taken on 2019 July 1 in C(4-8 GHz) and X(8-12 GHz) bands, in the BnA configuration. A broadband setup, ultimately covering the whole 8 GHz bandwidth across both bands was used. The target was observed for 15.5~minutes in each band. We observed 3C286 as the flux density scale and bandpass calibrator, and J1224+0330 as the gain calibrator. For polarization calibration 3C286 was used as the polarization angle calibrator, and J1407+2827 as the leakage calibrator. These radio observations provide us with sub-arcsec angular resolution, and $\mu$Jy level image sensitivity.

In order to investigate the unusual radio morphology of \srcname, observations of the southwestern radio structure of the target were also obtained with the Very Long Baseline Array (VLBA), with the addition of a single VLA antenna (VLBA+Y1) to provide short-spacing baselines, on 2022 Mar 25. A dual polarization setup was used to observe the target source for 28 minutes in C-band (4.1-4.5 GHz). We observed 4C39.25 as the fringe finder, J1231+0418 and J1239+0443 as the phase-reference calibrators, and 1219+044 as an amplitude check source. These long-baseline observations provide us with milliarcsec angular resolution, and $\mu$Jy level image sensitivity.

\subsubsection{VLA Total Intensity Calibration and Imaging}\label{sec:RadioCalImg}
Calibration of the radio data was done with the Common Astronomy Software Application (CASA). For the VLA-only data, we used its standard calibration pipeline (CASA 6.2.1.7). The calibrated target data were then inspected by eye to determine if any additional flagging of left over radio frequency interference (RFI) was needed. In total 45.7\% of the data were flagged.

Imaging of the source was done using CASA's \textit{tclean} task with a standard gridder and a multi-scale multi-frequency synthesis (mtmfs) deconvolver. We used two Taylor terms during deconvolution, corresponding to a power law, to enable modeling of the spectral properties within the large bandwidths being imaged. Due to the asymmetric nature of the BnA configuration point-spread-function (PSF) for these observations, Briggs weighting with a robust parameter of 1.0 was implemented to produce a more symmetrical PSF for deconvolution. This weighting also favors the shorter baseline signals and thus enhances the sensitivity to extended emission. A number of images, with different bandwidths, were created, as follows. First, in order to study the SED of the source across the C and X bands, we imaged both bands in 1~GHz bandwidth chunks. We repeated this imaging of 1~GHz bandwidth chunks with an added restriction to the $uv$ range of the data such that the largest angular scale is consistent across the images. These $uv$-restricted images are only used to extract the integrated radio SED,  which ensures that we are not introducing a systematic bias into our measured SED. The angular resolution for each 1~GHz subimage was matched to the lowest angular resolution as given by the 4--5~GHz subimage, $1.6\times 0.5$ arcsec$^2$. The rms levels for the flat-noise subimages vary from 12.4 ~$\mu$Jy/beam (4.5 GHz) to 24.0~$\mu$Jy/beam (11.5 GHz). The rms  increases with frequency except for 6.5 and 7.5 GHz, which have rms = 12.1 and 11.9 $\mu$Jy/beam, respectively. The rms levels are the same for the $uv$-restricted and unrestricted images. 

Larger bandwidth images were required for the detection of the faint radio core at higher frequencies; we created a 3~GHz bandwidth image at 10.5~GHz for this purpose. The image centered at 10.5~GHz was smoothed to the resolution of the 4.5~GHz image ($1.6\times0.5$ arcsec$^2$) and the rms levels in this image was 11.0~$\mu$Jy/beam. All images were primary beam corrected to allow for the correct interpretation of the extracted SED's spectral properties.

\subsubsection{VLA Polarization Calibration and Imaging}
The CASA standard calibration pipeline by default applies parallactic angle corrections without performing polarization calibration steps. Therefore, we removed the correction and recomputed the data weights before our manual calibration of the polarization products. The data were then further inspected for RFI. The instrumental (multiband) delays between the cross-hands were derived using 3C286, and the leakage terms were derived using J1407+2827 as the unpolarized calibrator. The data were then flagged a final time to remove residual RFI revealed by these calibration steps. In total, 47.5\% of the full 4-12 GHz data was flagged. The resulting data were then imaged in full Stokes (IQUV) across the full 4-12 GHz bandwidth, again with standard gridding and an mtmfs deconvolver with two Taylor terms. Again, Briggs weighting with a robust parameter of 1.0 was used.  We also imaged the data in full Stokes in 1 GHz bandwidth chunks in order to investigate how the polarized intensity and polarization position angle of the observed emission vary with wavelength.

\subsubsection{VLBA Total Intensity Calibration and Imaging}
For the VLBA+Y1 radio data, we followed the steps outlined in \citet{Linford2022}. Specifically, the data amplitude corrections were determined from autocorrelations between stations, system temperature and gain curve information were generated from the data, instrumental delays were determined as a function of frequency for all time, and then a global fringe fit was performed to determine delays as a function of frequency and time using 45s solution intervals. Next, the shape of the bandpasses were corrected using the fringe finder 4C39.25, and then  final amplitude scaling was applied. During the calibration process of the VLBA+Y1 data, 11.2\% of the data was flagged.

The VLBA+Y1 data were imaged using CASA's \textit{tclean} task with a standard gridder and an mtmfs deconvolver. We used two Taylor terms during deconvolution, corresponding to a power law, to enable modeling of the spectral properties within the large bandwidths being imaged. Natural weighting was used to favor shorter baselines and thus enhance the sensitivity to extended emission. The angular resolution and rms of the resulting image are 9.05 $\times$ 2.38 miliarcsec$^2$ and 23.1 $\mu$Jy/beam, respectively. 

 \subsubsection{Supplementary Radio Data}

In our analysis we refer to images from the  VLA Faint Images of the Radio Sky at Twenty-Centimeters \citep[FIRST; ][]{Becker1994} and VLA Sky Survey \citep[VLASS;][]{Lacy2020} radio surveys. The FIRST survey has an angular resolution of 5.4\arcsec and an rms level of 0.14~mJy/beam at 1.4 GHz. The VLASS survey is ongoing, and at present we use its quick look images with angular resolutions of $2.5\times2.1$ arcsec$^2$ and $2.5\times2.0$ arcsec$^2$ for epochs one and two, respectively. Stacking images from these epochs resulted in a noise level of 0.11 mJy/beam at 3~GHz.  We also include low-frequency observations of \srcname\ from the LOFAR Virgo Cluster survey \citep{Edler2023}, which contains \srcname\ in the background of the targeted cluster, and auxiliary low-frequency low-resolution measurements from various radio sky surveys. The LOFAR Virgo cluster survey has an angular resolution of $9.0\times5.0$ arcsec$^2$ and median rms level of 0.28 mJy/beam at 120--168 MHz. The low resolution surveys covering the location of \srcname\ span radio frequencies only up to 1.4~GHz and are incompatible with our high resolution data due to the latter resolving out diffuse emission on angular scales $>10-15$~arcsec. Therefore these data are treated separately (Table~\ref{tab:LownuFluxes}).

\begin{deluxetable*}{ccl}[t!]
\caption{Total radio flux density measurements from low resolution radio surveys. The specific surveys are: GLEAM -- Galactic and Extragalactic All-sky MWA (Murchison Widefiled Array) survey, VICTORIA -- Virgo Cluster multi-telescope observations in radio of interacting galaxies and AGN, TGSS -- TIFR (Tata Institute of Fundamental Research) GMRT (Giant Metrewave Radio Telescope) Sky Survey, RACS -- Rapid ASKAP (Australian Square Kilometre Array Pathfinder) Continuum Survey, NVSS -- NRAO VLA Sky Survey. The measurements were taken from the surveys' catalogs unless stated otherwise.}
\label{tab:LownuFluxes}
\centering
\tablewidth{0pt}
\tablehead{\colhead{Frequency} & \colhead{S$_{\nu}$ [mJy]} & \colhead{Notes and references}}
\startdata
 88 MHz & $252\pm56$ & GLEAM$^a$, \citet{HurleyWalker2017}\\ 
118 MHz & $260\pm30$ & GLEAM$^a$, \citet{HurleyWalker2017}\\ 
144 MHz & $221\pm44$ & VICTORIA$^b$, \citet{Edler2023}\\
150 MHz & $155\pm13$ & TGSS$^c$, \citet{Intema2017}\\ 
154 MHz & $194\pm22$ & GLEAM$^a$, \citet{HurleyWalker2017}\\ 
200 MHz & $189\pm26$ & GLEAM$^a$, \citet{HurleyWalker2017}\\ 
887 MHz & $65.8\pm5.3$  & RACS, \citet{Hale2021}\\ 
1.37 GHz & $48.1\pm0.8$ & RACS, \citet{Duchesne2023}\\ 
1.40 GHz & $42.3\pm1.7$ & NVSS, \citet{Condon1998}\\ 
\enddata
\tablenotetext{ }{(a) Measured from 30~MHz bandwidth image. (b) Measured from the image. (c) Lower flux density from the TGSS~ARD1 (Alternative Data Release 1) catalog as compared to VICTORIA and GLEAM may be due to its known calibration issues.}
\end{deluxetable*}

\subsection{Optical Spectroscopic Observations}
\begin{table*}[t!]
\centering
\begin{center}
\caption{APO Observations of \srcname}
\label{tab:observations}
\begin{tabular}{lllllllll}
\hline
\hline
{ Instrument} & \# Exps & PA & Slit size & Spectral Resolution & Exposure Time & Seeing & Airmass & UT Date \\ \hline
{ DIS} & 4   & 24.2$^{\circ}$  &   1.5\arcsec  &   R$\sim$3200-3900             & 600           &  2.45\arcsec      & 2.02    &   03/06/2021       \\
{ KOSMOS} & 5   & 22.5$^{\circ}$  &   1.2\arcsec &      R$\sim$2300            & 1200           & 1.67\arcsec   & 1.47   &    04/21/2023  \\   
\hline
\hline
\end{tabular}
\end{center}
\end{table*}
Spectroscopic observation of \srcname\ were conducted at APO on the 3.5m telescope using the DIS long-slit spectrograph on the night of UT 2021 March 6, and using the KOSMOS long-slit spectrograph on the night of UT 2023 April 21. The dual imaging feature of DIS allows simultaneous observations of the blue (2570--6230 \AA) and red (5190--9810 \AA) wavelength ranges, and allows for a user-specified central wavelength in each range. We use DIS to obtain medium resolution (R$\sim$3500) spectra of \srcname\ centered on the expected locations of the [O~III] doublet and the [O~II] doublet on the red and blue sides, respectively. KOSMOS allows for a user-specified combination of slit size, slit location, and disperser (red or blue) to control the wavelength range and spectral resolution. We use KOSMOS to target other emission lines of interest on the red side of the \srcname\ spectrum (H$\alpha$, [N~II]$\lambda\lambda$6548,6583, [S~II]$\lambda\lambda$6716,6730), while still retaining moderate spectral resolution (R$\sim$2000).

The DIS observations were obtained with a position angle aligned to the radio jet axis of the source (24.2$^{\circ}$), as inferred from the alignment of the two hotspots. The pixel scales for DIS are 0.42\arcsec/pixel for the blue side and 0.40\arcsec/pixel for the red side. The DIS observations were taken with central wavelengths 4798\AA\ and 6600\AA\ respectively, as well as a 1.5" slit, resulting in a resolution element of 1.5-1.7 \AA\ as determined by a bright skyline near the central wavelengths. The resolving power of these DIS observations was R$\sim$3200-3900. Four exposures of 600s were obtained with this setup. In these DIS observations we identified evidence of a detached cloud to the South of \srcname\ along the direction to the southern radio structure. This motivated us to shift the KOSMOS position angle to align with the radio core and the brightest part of the southern radio structure. The KOSMOS observations were obtained with a position angle of (22.5$^{\circ}$). The pixel scale for KOSMOS is 0.258\arcsec/pixel. The central wavelength of the KOSMOS spectrum, which was taken with a 1.18~\arcsec\ slit, was 7280 \AA, resulting in a resolution element of 3.15 \AA\  as determined by a bright skyline near the central wavelengths, and a resolving power of R$\sim$2300. Five exposures of 1200s were obtained with this setup. Details of these observations, including seeing, position angles, and airmasses are listed in Table \ref{tab:observations}.

\begin{figure*}[ht!]
    \centering
    \includegraphics[width=\linewidth]{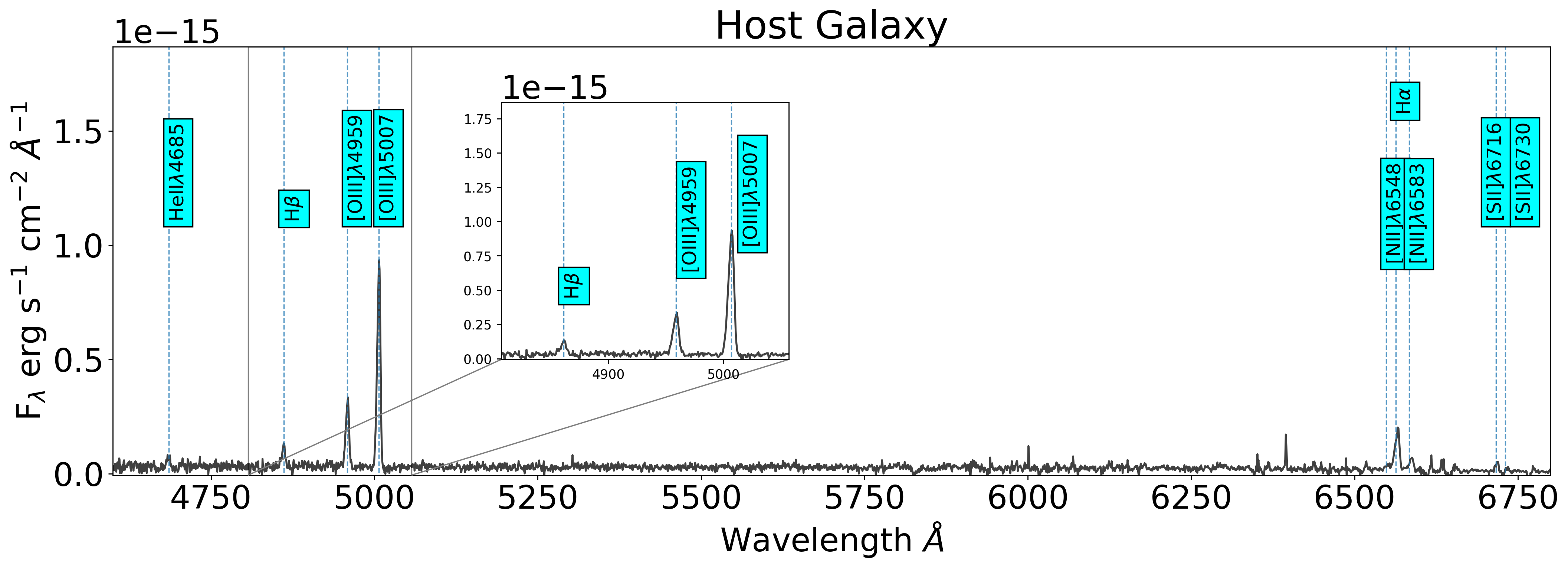}
    \includegraphics[width=\linewidth]{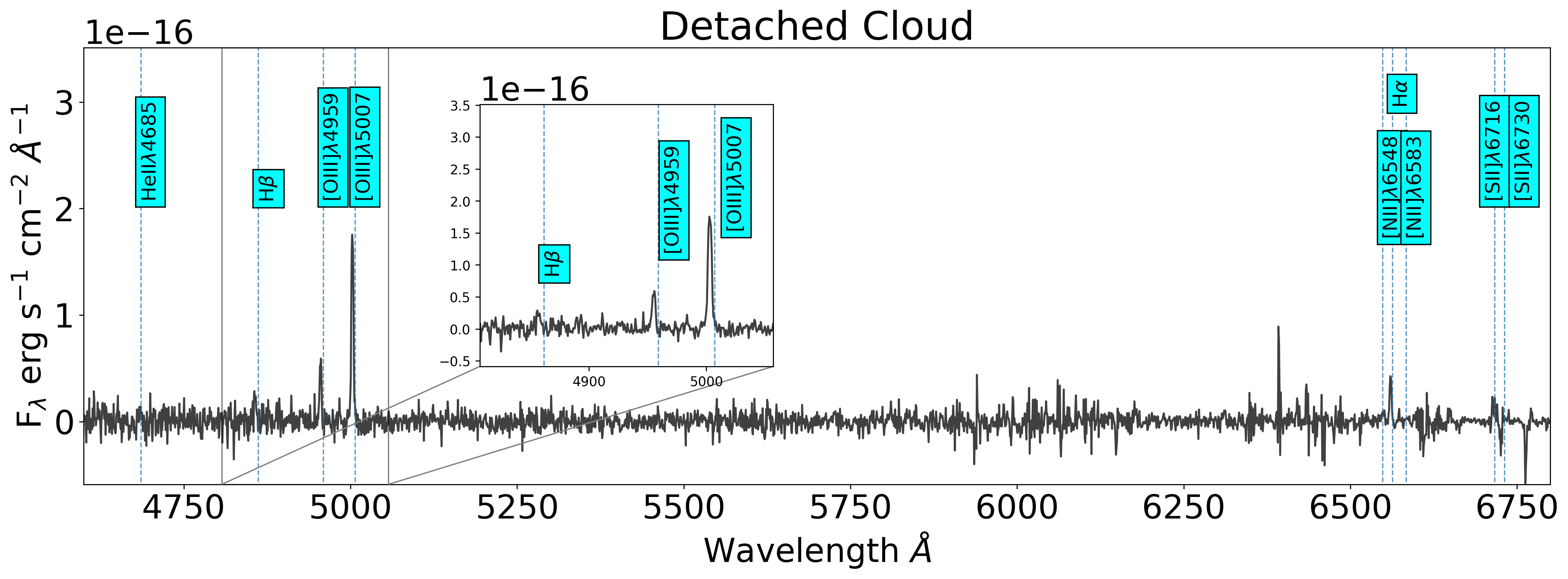}
    \caption{Top: Low resolution optical APO KOSMOS spectrum for the core of \srcname\, extracted from a 8.8\arcsec\ wide aperture. Bottom: Low resolution optical APO KOSMOS spectrum for the detatched cloud in \srcname's spectrum, extracted from a 5.0\arcsec\ wide aperture. In both spectra, the KOSMOS long-slit is aligned with the radio core and the brightest part of the southern radio component. The [OIII] line is the brightest feature in these spectra.}
    \label{fig:OptSpec_KOSMOS}
\end{figure*}

\begin{figure*}[ht!]
    \centering
    \includegraphics[width=\linewidth]{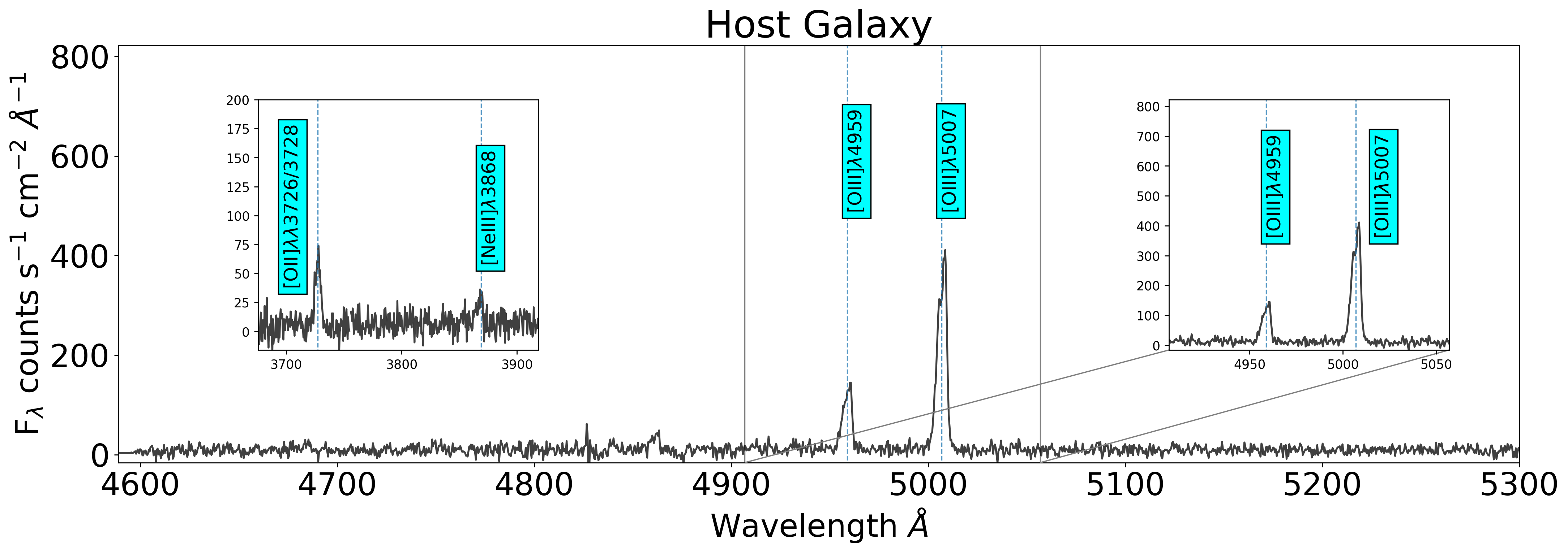}
    \includegraphics[width=\linewidth]{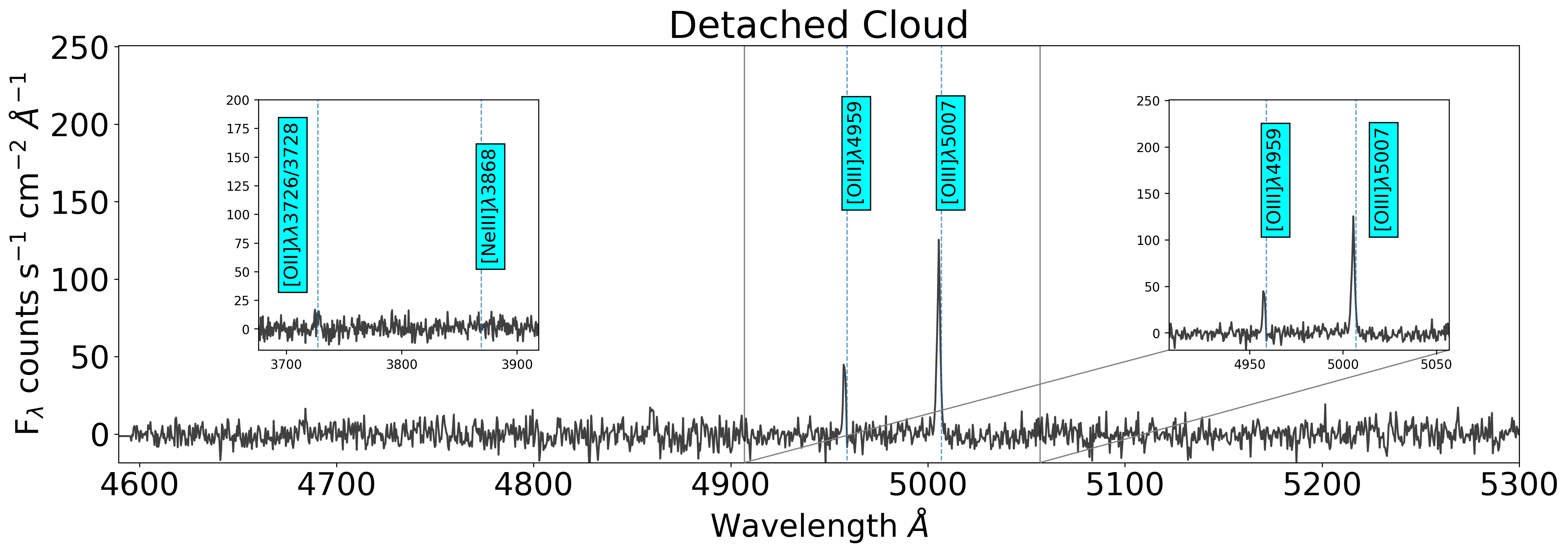}
    \caption{Top: Medium resolution optical APO DIS spectrum for the core of \srcname\, extracted from a 8.8\arcsec\ wide aperture. Bottom: Medium resolution optical APO DIS spectrum for the detatched cloud in \srcname's spectrum, extracted from a 5.0\arcsec\ wide aperture. In both spectra, the DIS long-slit is aligned with the radio jet axis of the source as determined from the location of the NE and SW hotspots. The [OIII] doublet line profiles of the host galaxy (top panel) suggests multiple kinematic components are present, and the [OIII] doublet line profiles of the cloud appear narrow, symmetric, and slightly blue shifted with respect to the central wavelengths of the emission lines in the host galaxy spectrum.}
    \label{fig:OptSpec_DIS}
\end{figure*}

\subsubsection{Data Reduction}\label{sec:OptDataRed}
The reduction of the APO observations were carried out using Pyvista \footnote{https://github.com/holtzmanjon/pyvista}, a python-based reduction software written specifically for the standard reduction of various data products produced at APO. For each set of observations, individual exposures were first bias-subtracted, flat-fielded, and then median-combined into a 2-D spectrum. A 2-D wavelength solution determines the wavelength solutions in each row of the master images, where rows are different spatial locations along the slit. The 2-D wavelength solution was found using HeNeAr calibration lamp exposures taken at the beginning of the night. The resulting rms accuracy for both the blue and the red channels in the 2-D wavelength solution for DIS medium-resolution spectra was 0.02 \AA, and the resulting rms accuracy of the 2-D wavelength solution for the KOSMOS observations was 0.07 \AA. The 2-D wavelength solutions were then used to correct for deviations in wavelength solution along the spatial axis of the target 2-D spectra by interpolating the spectra on a row-by-row basis to the wavelengths at the location of the target trace. This ensures that the wavelengths are constant along the spatial direction and do not drift with distance from the target trace due to e.g. flexure issues. For observations of the red side of \srcname's KOSMOS spectrum, the final wavelength solution at the location of the target trace was corrected using the observed optical skylines, with a resulting rms accuracy of 0.15 \AA. Due to  the lack of  standard star  observations on the night of UT 2021 March 6, we omit flux calibrating our medium-resolution DIS data. The KOSMOS data were flux calibrated using standard star observations of Feige98 and corrected for Galactic extinction using the E(B-V) values from the dust maps of \citet{Schlegel1998}, the standard extinction curve from \citet{Fitzpatrick1999}, and R$_{V}$=3.1. 

The sizes of the slits in both data sets (1.2-1.5\arcsec; Table~\ref{tab:observations}) are narrower than the width of \srcname\ on the sky, and narrower than the seeing during the observations, which means that some of the source's light will be missed. This loss of light should not affect the measured line central wavelengths, line widths, or ratios between different lines, however, this does mean that our measured fluxes are only lower limits to the total source fluxes. We do not perform slit-loss corrections to account for these losses. 

Both 2-D target spectra revealed that the extended emission ([O~II]$\lambda$3726,3728, [Ne~III]$\lambda$3868, [O~III]$\lambda$5007,4959, H$\alpha$) in \srcname\ is fractured along the radio axis, with the host galaxy component and associated continuum, and a detached cloud to the south-west of the host galaxy. One-dimensional spectra were extracted from the 2-D spectra as follows. First, we extracted a spectrum for the host galaxy alone using a 8.8\arcsec\ wide aperture. Then, we extracted a spectrum for the detached cloud alone using a 5.0\arcsec\ wide aperture. In both of the rest-frame KOSMOS and DIS spectra, shown in Figures~\ref{fig:OptSpec_KOSMOS} and \ref{fig:OptSpec_DIS}, respectively, we also include labels indicating key emission lines. An inset displaying the blue channel of the DIS spectrum, highlighting the [O~II]$\lambda$3726,3728 emission line detection in the core, is also included. The inset to the right side of the [O~III]$\lambda\lambda$4959,5007 doublet in each panel in Figures~\ref{fig:OptSpec_KOSMOS} and \ref{fig:OptSpec_DIS} displays a zoom-in on the [O~III] doublet, highlighting the line profiles of the [O~III] doublet.

\subsubsection{Fitting Emission Lines}
\begin{deluxetable*}{ccccccc}[htb!]
\caption{Observed emission line properties from the APO KOSMOS observations of \srcname\, separated by component. The host galaxy measurements were taken using a 8.8\arcsec\ wide aperture centered on the spatial centroid of the continuum. The cloud measurements were taken using a 5.0\arcsec\ wide aperture centered on the spatial centroid of the [O III]$\lambda$5007 emission.}
\label{tab:LineInfoKOSMOS}
\centering
\tablewidth{0pt}
\tablehead{ \colhead{Component} & \colhead{Line} & \colhead{Flux} & \colhead{Equivalent} & \colhead{Rest frame $\lambda$} \\ 
& & & \colhead{width} & }
\startdata
& & 10$^{-15}$[erg s$^{-1}$cm$^{-2}$] & [\AA] & [\AA] \\
\hline
Host Galaxy & He II$\lambda$4685 & 0.304 $\pm$ 0.051 & -10.396 $\pm$ 2.051 & 4684.764\\
& H$\beta$ & 0.572 $\pm$ 0.074 & -18.597 $\pm$ 3.432 & 4861.162 \\
& [O III]$\lambda$4959 & 1.837 $\pm$ 0.055 & -57.688 $\pm$ 4.055 & 4958.831 \\
& [O III]$\lambda$5007 & 5.469 $\pm$ 0.054 & -211.088 $\pm$ 14.293 & 5006.877 \\
& [N II]$\lambda$6548 & 0.205 $\pm$ 0.038 & -14.669 $\pm$ 6.645 & 6550.61 \\
& H$\alpha$ & 1.511 $\pm$ 0.048 & -109.806 $\pm$ 22.489 & 6565.379 \\
& [N II]$\lambda$6583 & 0.615 $\pm$ 0.114 & -47.52 $\pm$ 26.369 & 6586.02 \\
& [S II]$\lambda$6716 & 0.224 $\pm$ 0.033 & -24.971 $\pm$ 4.582 & 6718.152 \\
& [S II]$\lambda$6730 & 0.077 $\pm$ 0.019 & -10.063 $\pm$ 2.742 & 6732.522 \\
\hline
Cloud &  He II$\lambda$4685 &  0.051 $\pm$  0.031 & --  &  4687.934 \\
&  H$\beta$ &  0.108 $\pm$  0.027 & --  &  4860.685 \\
& [O III]$\lambda$4959 & 0.208 $\pm$ 0.022 & --  & 4958.991 \\
& [O III]$\lambda$5007 & 0.733 $\pm$ 0.025 & --  & 5006.816 \\
& H$\alpha$ & 0.159 $\pm$ 0.011 & --  & 6565.391 \\

\enddata
\end{deluxetable*}

\begin{deluxetable*}{cccc}[htb!]
\caption{Observed emission line location and velocity dispersion from the medium-resolution APO DIS observations of \srcname\, separated by component. The host galaxy measurements were taken using a 8.8\arcsec wide aperture centered on the spatial centroid of the continuum. The cloud measurements were taken using a 5.0\arcsec wide aperture centered on the spatial centroid of the [O III]$\lambda\lambda$4959,5007 emission.}
\label{tab:LineInfoDIS}
\centering
\tablewidth{0pt}
\tablehead{ \colhead{Component} & \colhead{Line} & \colhead{Rest frame $\lambda$} & \colhead{$\sigma_{v}$}}
\startdata

& & [\AA] & [km s$^{-1}$] \\ 
\hline
Host Galaxy & [O II]$\lambda\lambda$3726,3728 & 3727.456 & 105.499 \\
& H$\beta$ & 4861.561 & 70.71 \\
& [O III]$\lambda$4959 & 4959.028 & 76.514 \\
& [O III]$\lambda$5007 & 5007.156 & 77.846 \\
\hline
Cloud & [O III]$\lambda$4959 & 4957.186 & 7.591 \\
& [O III]$\lambda$5007 & 5005.051 & 22.24 \\

\enddata
\end{deluxetable*}
We derive flux estimates for the emission lines ( He~II$\lambda$4685, H$\beta$, [O~III]$\lambda\lambda$4959,5007, H$\alpha$+[N~II]$\lambda\lambda$6548,6584, and [S~II]$\lambda\lambda$6717,6731) seen in the KOSMOS observations (Figure~\ref{fig:OptSpec_KOSMOS}) by fitting a Gaussian to each line. The [O~III]$\lambda\lambda$4959,5007 doublet in the KOSMOS host galaxy spectrum (Figure~\ref{fig:OptSpec_KOSMOS}; Top panel) is used to constrain the redshift of the host galaxy to be 0.304458 $\pm$ 0.000007, which is consistent to within the SDSS photometric estimate error ($z=0.269\pm0.058$). This redshift is used to determine the central wavelength of the other emission lines for the host galaxy and the detached cloud in Table~\ref{tab:LineInfoKOSMOS}. 

For individual lines that are well separated in the KOSMOS data (He~II$\lambda$4685, [O~III]$\lambda$4959, [O~III]$\lambda$5007, and H$\beta$) we fit each line with a single Gaussian profile. For the H$\alpha$+[N~II]$\lambda\lambda$6548,6584 complex all three lines were fit simultaneously, requiring that each individual line in the [N~II] doublet has the same width, and that the [N~II]$\lambda\lambda$6548,6584 has the theoretical flux ratio 3.0 \citep{Storey2000,Ivan2023}. For the [S~II]$\lambda\lambda$6717,6731 doublet, both lines were fit simultaneously, requiring that each individual line in the [S~II] doublet has the same width.

We use these fits to measure the flux in each line, and we summarize these measurements in Table~\ref{tab:LineInfoKOSMOS}, along with the equivalent width of each line. For the cloud, which does not have a continuum detected above the noise in the spectra, we do not report the equivalent width.  The signal in the [S~II]$\lambda$6730 line is relatively weak and we note that the [S~II]$\lambda\lambda$6716,6730 doublet may have been affected by an over-subtracted skyline.

The [O III]$\lambda\lambda$4959,5007 doublet in the DIS host galaxy spectrum (Figure~\ref{fig:OptSpec_DIS}; Top panel) gives a redshift of z=0.305184 $\pm$ 0.000011, which is used to determine the central wavelength corrected for the systemic redshift of [O~III] and line dispersion of the other emission lines for the host galaxy and the detached cloud in Table~\ref{tab:LineInfoDIS}. For each emission line ([O II]$\lambda\lambda$3726,3728, H$\beta$, [O III]$\lambda\lambda$4959,5007) in the DIS spectra shown in Figure~\ref{fig:OptSpec_DIS}, we fit a single Gaussian to the line to determine the observed central wavelength and line dispersion ($\sigma_{v}$).  We then correct the line dispersion for instrumental broadening, and correct the observed central wavelength and the corrected line dispersion for the systemic redshift of the source. The resulting best-fit parameters are shown in Table~\ref{tab:LineInfoDIS}.

\subsubsection{Supplementary Optical Data}

 In our analysis, we refer to existing Hubble Space Telescope (HST) broadband imaging covering the location of \srcname\ as part of the Gems of the Galaxy Zoo (Zoo Gems) project gap-filler HST program \citep{Keel2022}. Observations of the \srcname\ were taken with a total exposure time of 674s in the Wide Field Camera (WFC) F625W filter (center = 6323.81~\AA, FWHM = 1389.28~\AA), covering the [O III]$\lambda\lambda4959,5007$+H$\beta$ line complex at the redshift of \srcname. The WFC/F625W image was made science ready by the ACS pipeline, which performs a drizzle combination of individual exposures, and corrects for charge transfer efficiency (CTE) issues, which could  otherwise lead to underestimating the pixel image counts. We use this imaging to aid in our understanding of the EELR around \srcname, the morphology of the host galaxy, and the evolution of the central AGN. We do not attempt to separate out the continuum emission from the image, but we do comment on how the included continuum light affects our interpretations, where necessary.

\section{Results} \label{sec:results}

\subsection{ Radio Properties and SEDs}

\begin{figure}[t!]
    \centering
    \includegraphics[width=\linewidth]{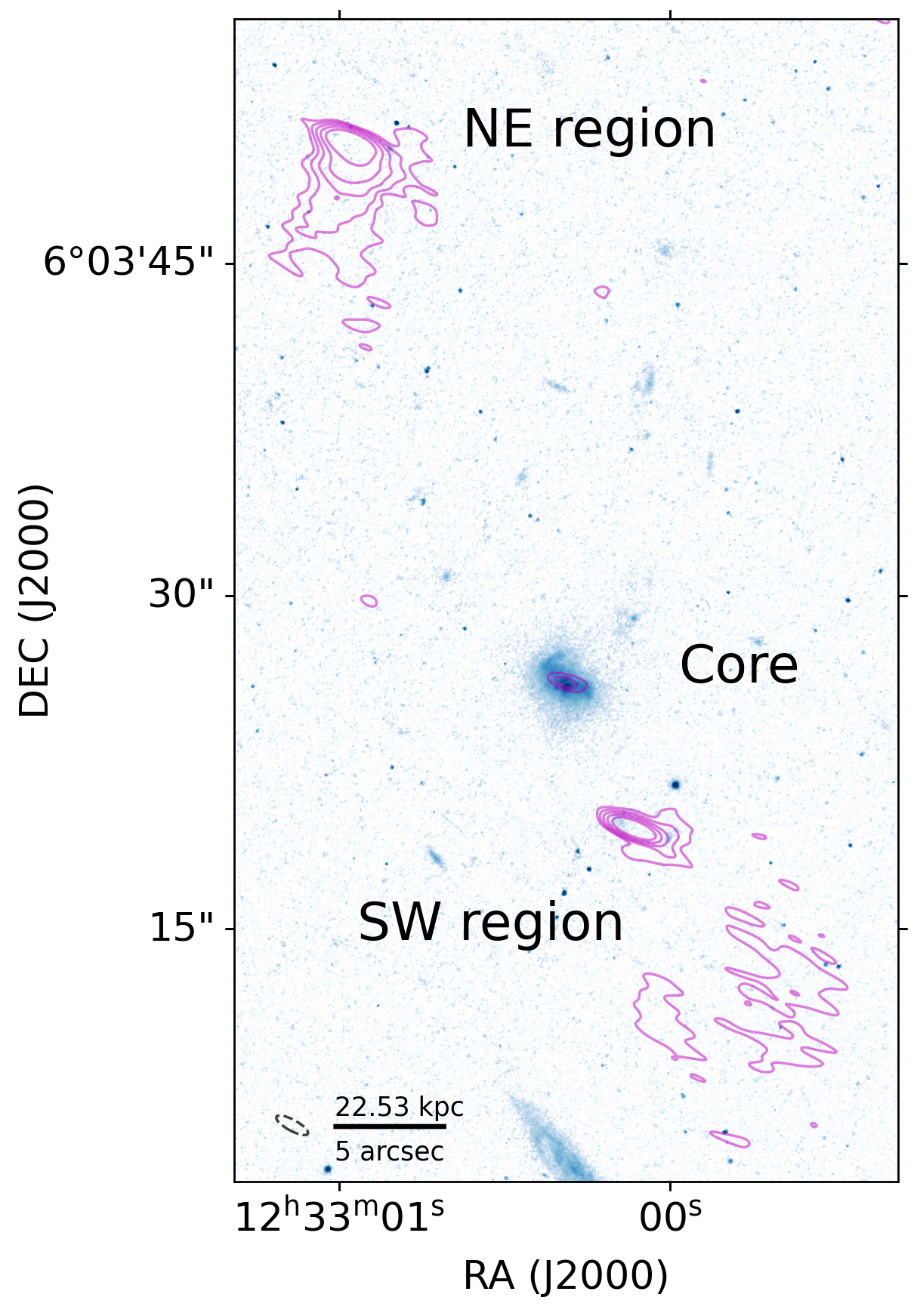}
    \caption{Radio and optical morphology.  HST F625W broadband optical image of \srcname\ with overlaid radio contours extracted from our 4.5~GHz radio image.  The color scale of the optical images is Log normalized between 0 and 1.  The radio contour levels are drawn down to 4$\sigma$ (50.8, 87.9, 151.9, 262.6, and 453.8 $\mu$Jy per beam). The beam size of the radio contours is $1.6\times 0.5$ arcsec$^2$, and is shown in the bottom left corner (dashed black) of each image. A scale bar indicating 5\arcsec\ is shown in the bottom left corner. The radio core of the emission is coincident with the optical host galaxy, confirming the association between the extended radio emission and the Green Bean Galaxy candidate host. } 
    \label{fig:rdoCtr_optIm}
\end{figure}

In Figure~\ref{fig:rdoCtr_optIm}, contours from the subimage containing the highest signal-to-noise detection of the radio source (S/N$\sim$~65 at a central frequency of 4.5~GHz) are overlaid on the  HST F625W broadband optical imaging of the host galaxy \citep{Keel2022}. These contours reveal three distinct radio components associated with this source: a northeast hotspot with diffuse lobe emission ( NE region, resolved), a southwest compact component with diffuse emission  trailing away from the radio core ( SW region, resolved), and an  unresolved compact core coincident with the optical host galaxy, \srcname.  The location of the  detected radio core unambiguously confirms the association of the radio galaxy with the GB host galaxy, originally suggested by \citet{Kapinska2017}.  Assuming that the two compact components are the terminal hotspots of the jets, the size of the radio galaxy is 33.3 arcsec (150 kpc) in projection (see Section~\ref{sec:morph} for further discussion on the radio morphology). The high resolution VLA data are not sensitive to angular scales larger than 10--15~arcsec in C band, hence we use LOFAR and NRAO VLA Sky Survey \citep[NVSS;][]{Condon1998} to measure total flux density of the radio source, which translates to radio luminosity densities of  $5.2(\pm1.1)\times10^{24}$~W/Hz\,sr at 144~MHz, and $1.0(\pm0.05)\times10^{24}$~W/Hz\,sr at 1.4~GHz.

\begin{deluxetable*}{cccc}[t!]
\caption{
Radio flux density measurements (S$_{\nu}$) centered on the compact components of the NE  region, SW region, and radio core of \srcname. All measurements were taken with the CASA's {\it imfit()} task, fitting a single Gaussian. The $uv$-range of each subimage was matched to that at the highest frequency of 12 GHz to avoid potential bias of emission being resolved out. The subimage bandwidth is 1~GHz unless stated otherwise}.
\label{tab:Fluxes}
\centering
\tablewidth{0pt}
\tablehead{\colhead{Frequency [GHz]} &  &\colhead{S$_{\nu}$ [mJy]} &  \\
\colhead{} & \colhead{NE region} & \colhead{SW region} &\colhead{Core}}
\startdata
4.5 & 3.78$\pm$0.09 & 1.61$\pm$0.04 & 0.117$\pm$0.019 \\
5.5 & 3.17$\pm$0.11 & 1.34$\pm$0.05 & 0.083$\pm$0.021 \\
6.5 & 2.85$\pm$0.12 & 1.18$\pm$0.05 & 0.076$\pm$0.020 \\
7.5 & 2.49$\pm$0.10 & 1.05$\pm$0.04 & 0.070$\pm$0.020 \\
8.5 & 2.37$\pm$0.11 & 0.88$\pm$0.05 & 0.060$\pm$0.016 \\
9.5 & 2.00$\pm$0.11 & 0.84$\pm$0.05 \\
10.5 & 1.85$\pm$0.13 & 0.77$\pm$0.06 & 0.047$\pm$0.016 $^a$ \\
11.5 & 1.55$\pm$0.11 & 0.60$\pm$0.05 

\enddata
\tablenotetext{} {(a) Measurement based on wideband (3~GHz) image with center frequency 10.5~GHz.}
\end{deluxetable*}

\begin{figure*}
    \centering
    \includegraphics[width=2.3in]{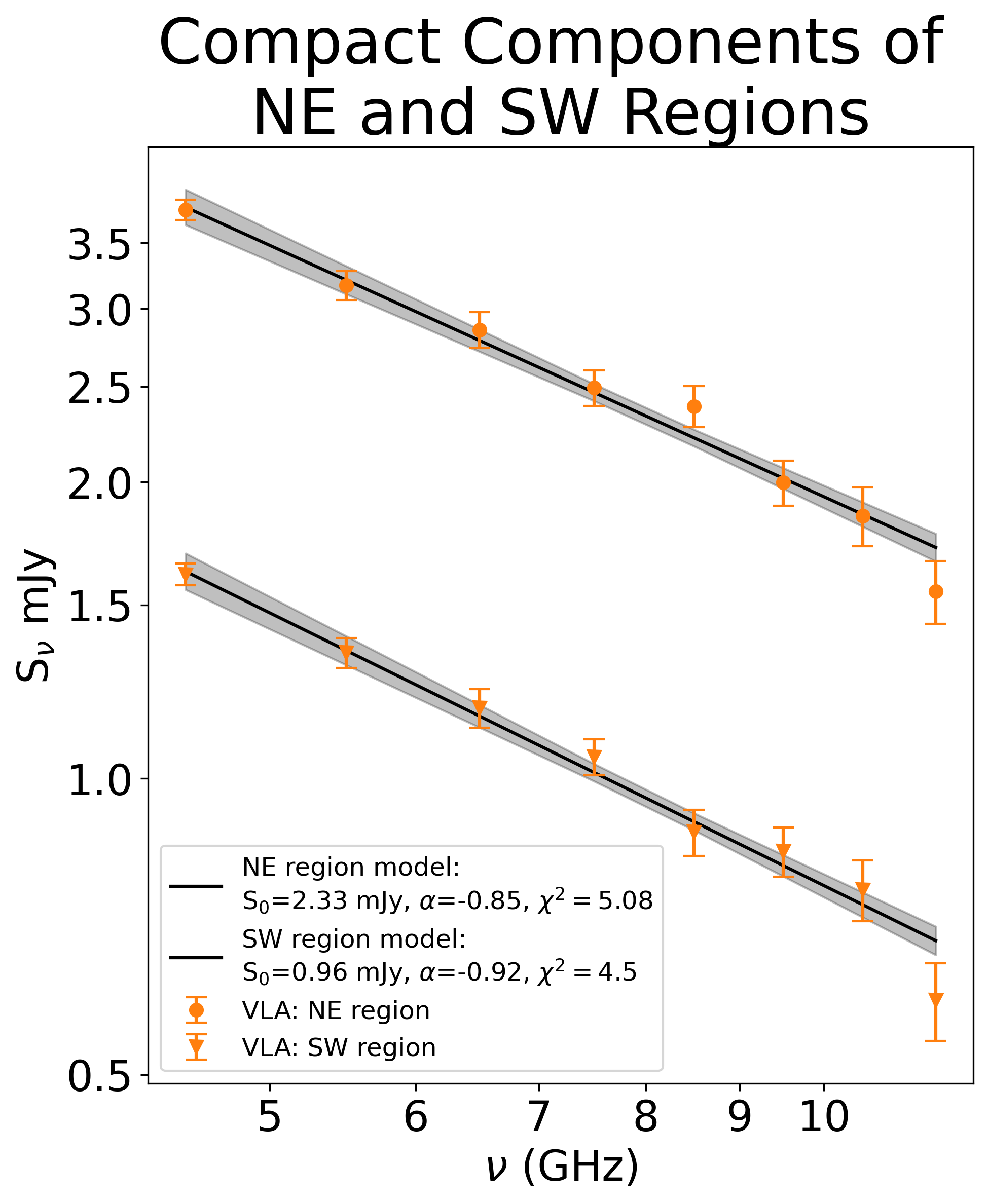}
    \includegraphics[width=2.3in]{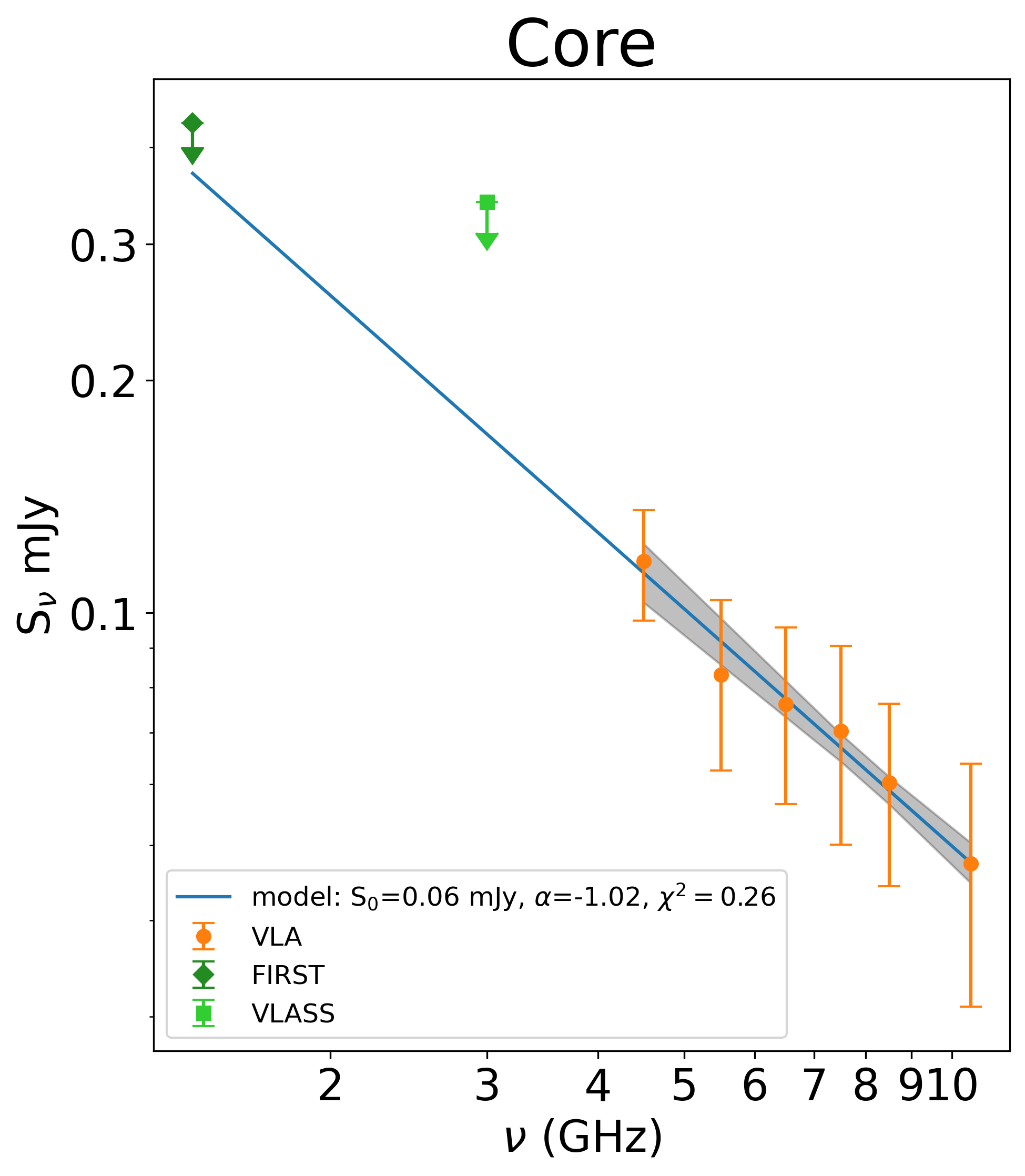}
    \includegraphics[width=2.3in]{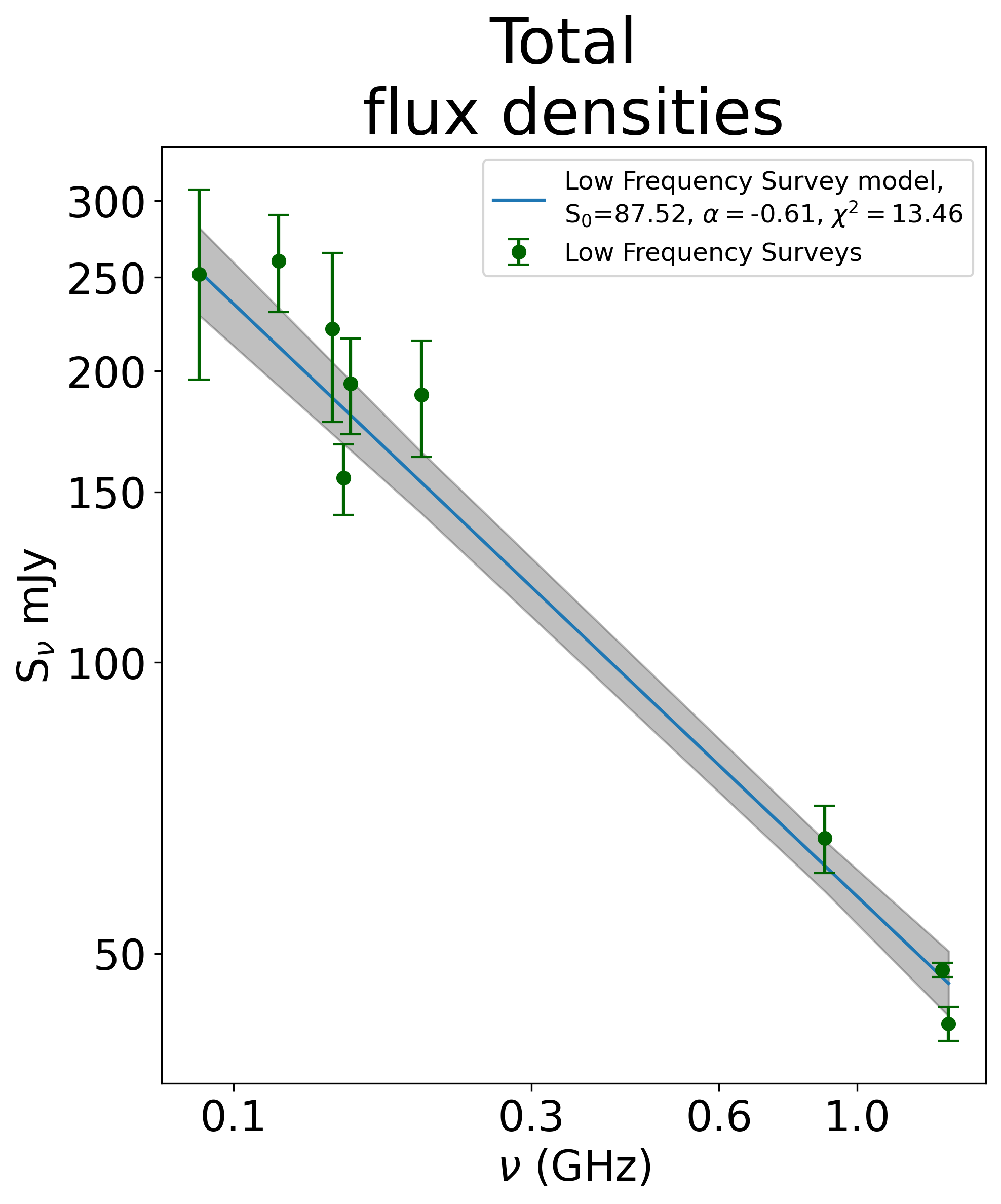}
    \caption{
    Radio SEDs of the  compact components of the NE  region (left), SW  region (middle), and Core (right). Our data presented in this paper are drawn as circles (orange). All three components were each best fitted by a simple power law with a reference frequency at 8.0~GHz. For the Core, FIRST and VLASS 3$\sigma$ upper limits are drawn as a diamond (dark green) and a square (light green), respectively. The upper limits on FIRST and VLASS measurements were not included in the fitting.}
    \label{fig:radioSEDs}
    \label{fig:NE_SED_p1GHz}
    \label{fig:lownuSED}
\end{figure*}

 The radio core is clearly detected in the 1~GHz bandwidth images created from our new higher sensitivity VLA observations up to 8.5 GHz, however, detection of the core at higher frequencies requires larger bandwidth images (Section~\ref{sec:RadioCalImg}). The integrated radio SED between 4.5~GHz and 11.5~GHz for each radio component are presented in Table~\ref{tab:Fluxes} and Figure~\ref{fig:radioSEDs}.  The components are fitted by  a single 2D Gaussian profile to the NE and SW regions, and the core.  We do not attempt to measure the low signal-to-noise diffuse emission in the SW region. The NE region measurements are resolved, and therefore we assume that these integrated flux densities include all of the hotspot emission, as well as some of the surrounding lobe emission.

The spectral index, $\alpha$, is defined as $S_\nu\propto\nu^{\alpha}$, where $S_\nu$ is the source flux density at  a frequency $\nu$. We find that the  compact radio components were each best fitted by a simple power-law between  4.5~GHz and 11.5~GHz with $\alpha^{11.5}_{4.5}=-0.85\pm0.04$  ($\chi^2=5.08$) for the NE  region, $\alpha^{11.5}_{4.5}=-0.92\pm0.04$  ($\chi^2=4.50$) for the SW  region, and $\alpha^{10.5}_{4.5}=-1.02\pm0.09$  ($\chi^2=0.26$) for the core, with reference frequency of 8.0~GHz.

The radio core  is undetected in the FIRST (dark green) and VLASS (light green) surveys with $3\sigma$ upper limits of 0.43~mJy/beam and 0.34~mJy/beam, respectively.  While the upper limits on the FIRST and VLASS measurements were not included in the fitting,   for reference we plot these data points along our high resolution VLA data (Figure~\ref{fig:lownuSED}).  We also show the total low-frequency flux densities up to 1.4~GHz in the last panel of Figure~\ref{fig:lownuSED}. We find that the total flux density of the source between 88~MHz and 1.4~GHz is best fitted by a simple power law, with $\alpha^{1.4}_{0.088}=-0.61\pm0.04$ ($\chi^2=13.46$) at a reference frequency of 200~MHz.

\subsection{Radio Spectral Analysis}
\label{sec:RadioSpectralAnalysis}
 To investigate these spectral shapes on resolved scales and estimate the spectral age of the radio-emitting plasma, 
we perform spectral analysis using the Broadband Radio Astronomy ToolS\footnote{http://www.askanastronomer.co.uk/brats} software package \citep[BRATS;][]{Harwood2013}. BRATS is a software for handling spectral analysis of broadband radio data, including calculations of resolved spectral index maps, fitting synchrotron spectral aging models to the observed data, and image reconstruction. We use the original untapered 1 GHz-wide images between 4.5 and 11.5 GHz as the input for BRATS. BRATS uses an `adaptive' method to identify the regions over which flux densities will be integrated and analyzed within a source image, with a set of limits defined by the user.  The data  are fitted on a pixel-by-pixel basis, and use the default parameters in BRATS \citep{Harwood2013}. Performing this adaptive source region identification resulted in a total of  616 regions, each 1 pixel$^2$  (0.1$\times$0.1~arcsec$^2$) in size, collectively covering  6.1 arcsec$^2$. The core of the radio emission was not included in the final BRATS source regions as spectral aging models to not apply to the radio emission produced in the nucleus.

BRATS allows the user to fit spectral aging models to the observed data on a region-by-region basis. For these models, the user can make some fitting choices to control the maximum source age to restrict the model fit (`{\it{myears}}' parameter), the number of times to decrease the range of ages to fit across and re-attempt the fit (`{\it{levels}}' parameter), and the resolution of each fitting attempt (`{\it{ageres}}' parameter). For our use of BRATS, we imposed a maximum source age of `{\it{myears}}'=35 Myears, the number of fitting levels as `{\it{levels}}'=3, and the resolution of each fitting attempt to be `{\it{ageres}}' = 10 increments. The user can then determine which model to use to fit to the data from the well-known \citet[][hereafter the `JP' model]{Jaffe1973}, \citet[][hereafter the `KP' model]{Kardashev1962}, and \citet[][]{Tribble1993} spectral aging models. In this work we use the JP spectral aging model which allows for a realistic description of the motions of particles, while still retaining a relatively simple description of the magnetic field. Specifically, the JP model assumes random/isotropic pitch angles while the KP model assumes a fixed pitch angle for all time. Both the JP and KP models assume a constant magnetic field.

The most important physical inputs when calculating the spectral aging model are the injection index ($\alpha_{inj}$), the magnetic field strength (B) of the radio source, and the redshift of the source. $\alpha_{inj}$ describes the initial distribution of electrons injected along the magnetic field lines at the point of jet disruption/termination. The magnetic field strength, B, determines the synchrotron radiative losses of the electron distribution. The redshift of the source determines the magnetic field strength of the Cosmic Microwave Background (B$_c$), and thus the inverse-Compton radiative losses of the electron distribution. Once a model had been created for a source, BRATS can then compute a reconstructed model image of the source, a spectral age map and associated errors, and a $\chi^2$ map to aid in assessing the goodness-of-fit across each region. 

\begin{figure*}[ht!]
    \centering
    \includegraphics[width=0.48\linewidth]{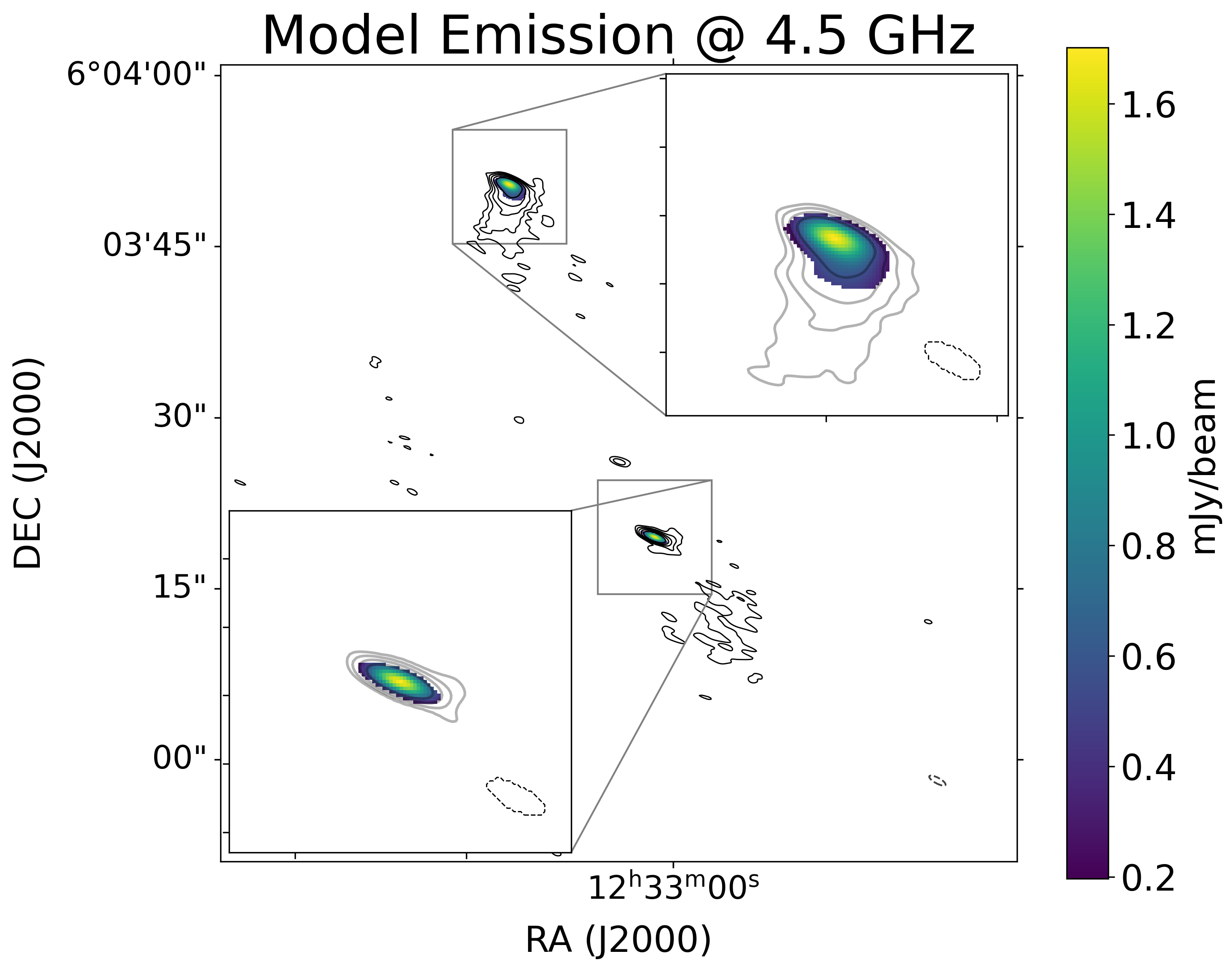}
    \includegraphics[width=0.5\linewidth]{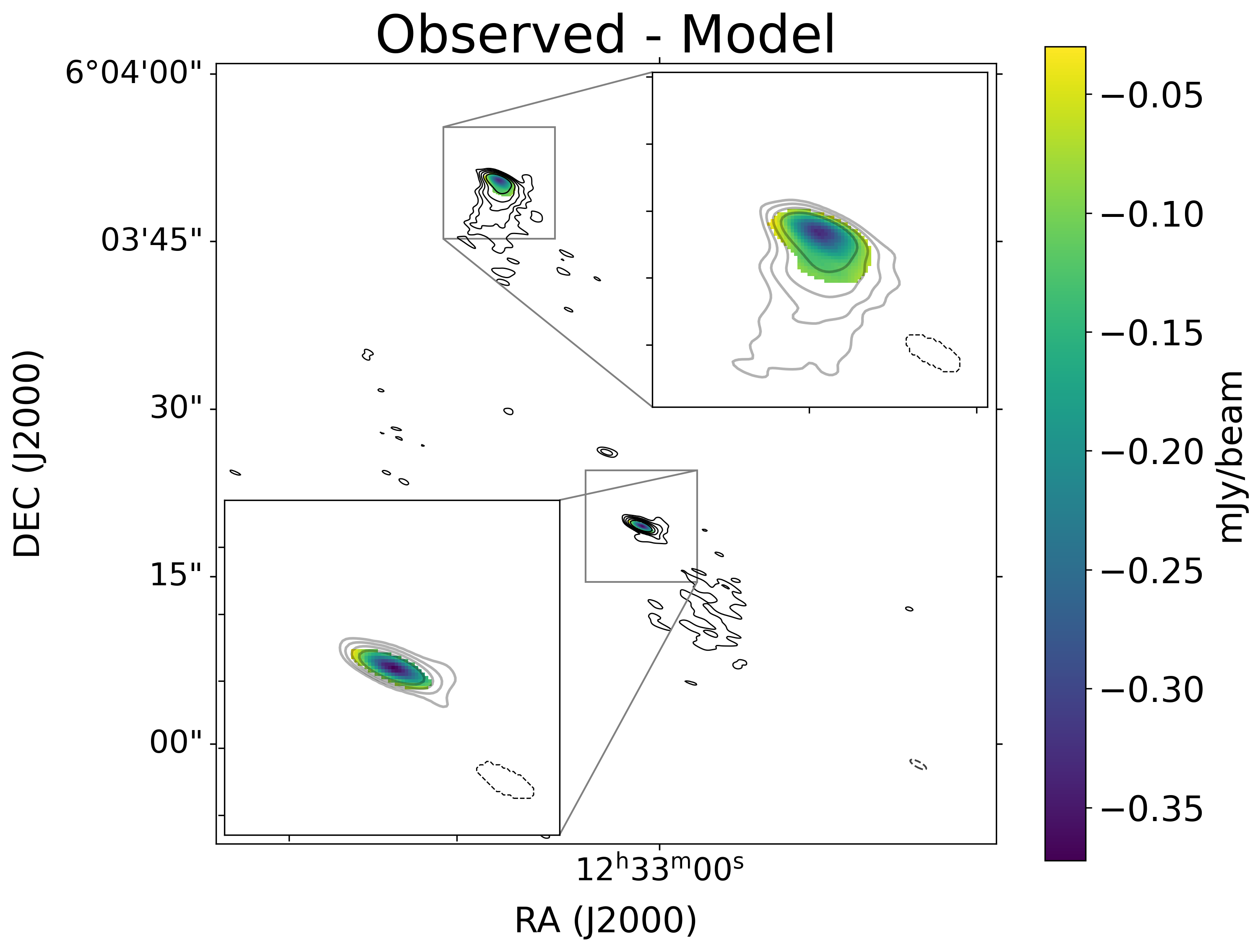}
    \caption{Left: Reconstructed model image from the best fitting spectral aging model. Right: Residual map showing the reconstructed modelled emission subtracted from the observed emission. VLA BnA configuration C band (4-5 GHz) contours down to 4.0$\sigma$ (50.8, 87.9, 151.9, 262.6, and 453.8~$\mu$Jy per beam) are included in each image. The beam size of the radio contours is $1.6\times 0.5$~arcsec$^2$, and is shown in the bottom right corner (dashed black)  of each image. The reader should note the difference in the the colorbar ranges for each panel. The largest residual (right panel) is $\sim-0.30$~mJy/beam.}
    \label{fig:FullSrc_reconmap}
    \label{fig:FullSrc_diffmap}
\end{figure*}

 For the magnetic field strength, we assume equipartition conditions. We use existing low-frequency imaging from the LOFAR Virgo cluster survey \citep{Edler2023} to measure a total source flux density of 221 $\pm$ 44 mJy at 150 MHz, indicating an effective magnetic field strength of B = 0.49 $\pm$ 0.03 nT \citep[assuming B=0.4$\times$B$_{eq}$;][]{Croston2005, Ineson2017}.

\begin{figure*}[ht!]
    \centering
    \includegraphics[width=0.6\linewidth]{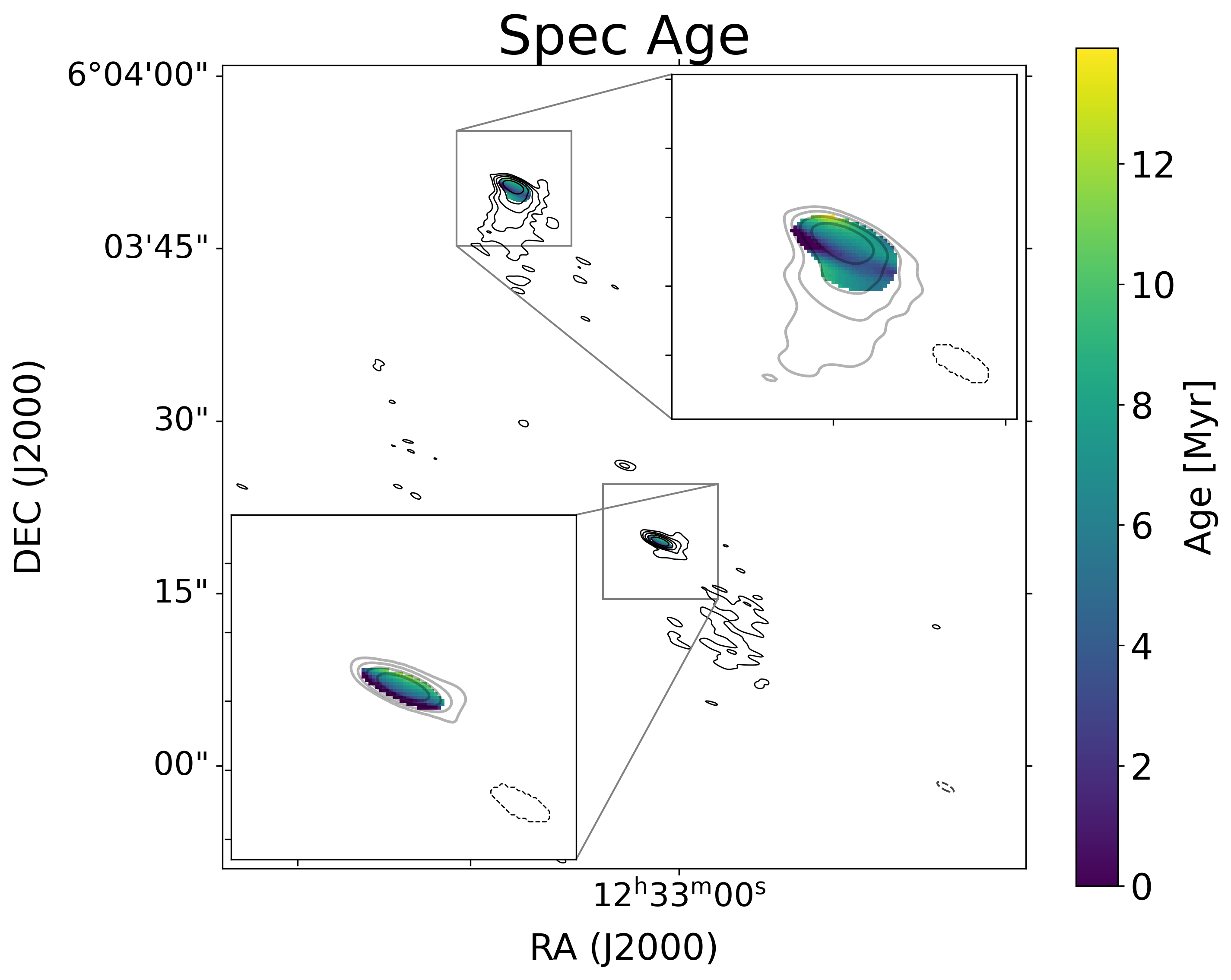}
    \includegraphics[width=0.47\linewidth]{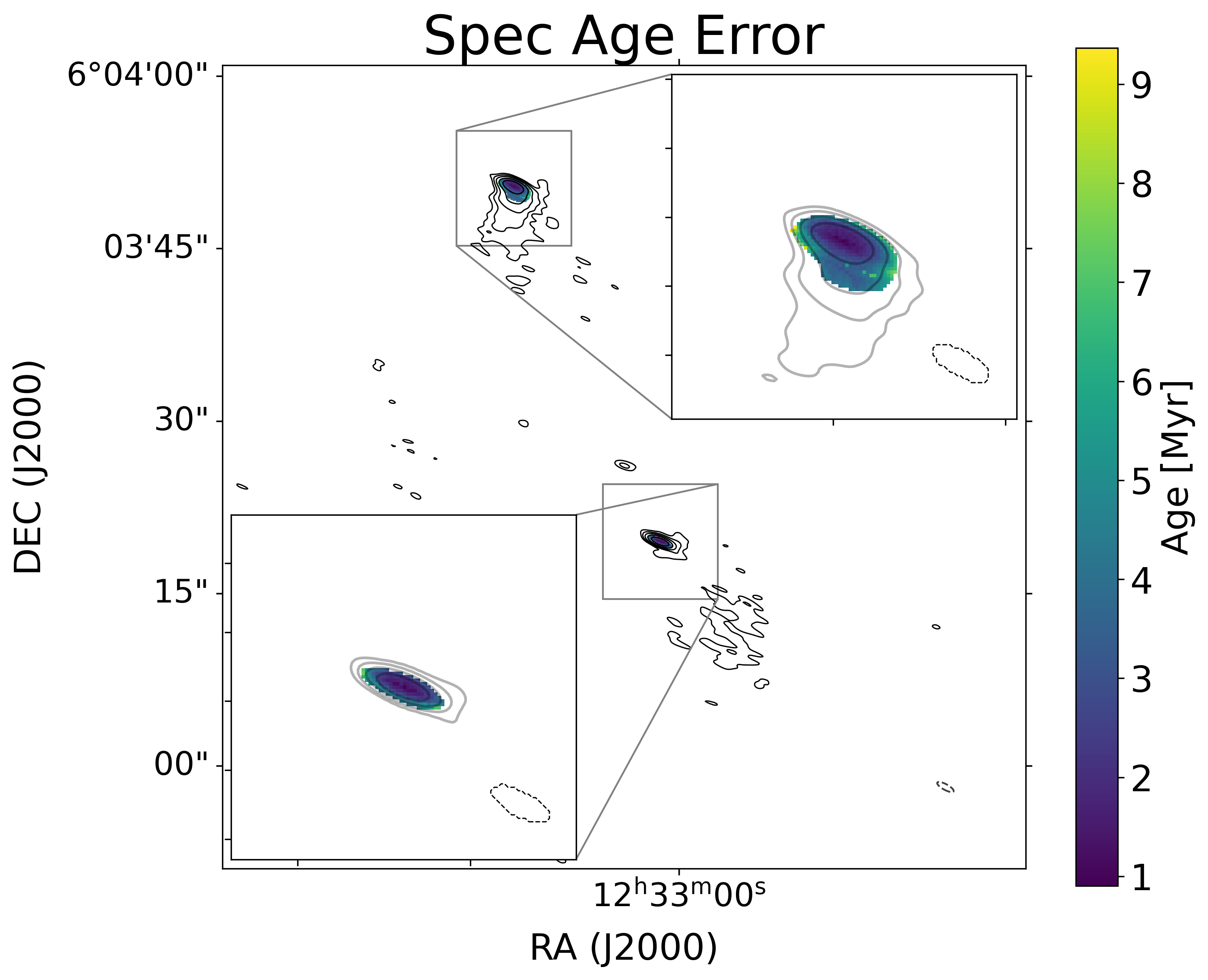}
    \includegraphics[width=0.49\linewidth]{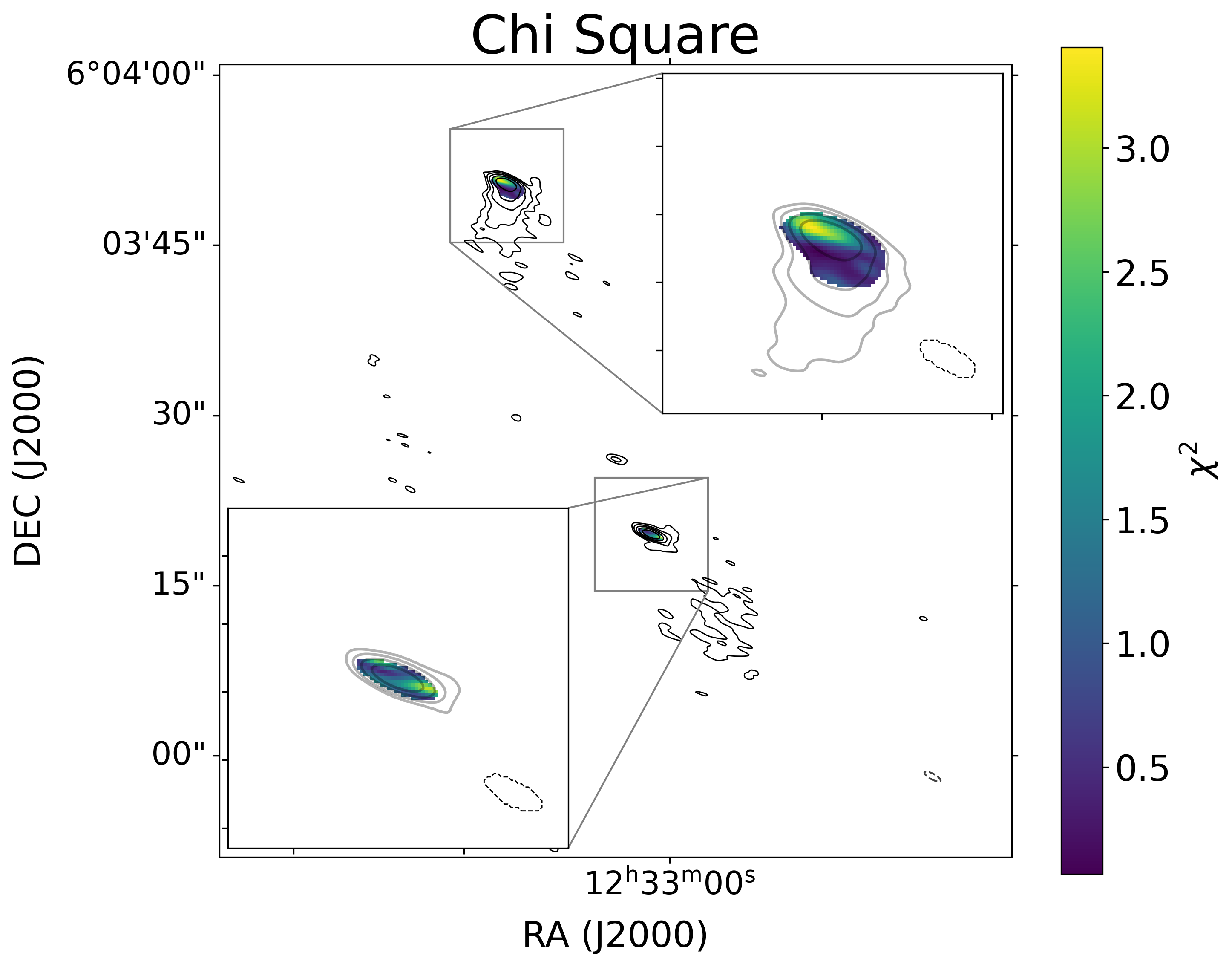}
    \caption{Top: Spectral Age map produced from the best fitting spectral age model run in BRATS.  Bottom left: Errors on the spectral ages shown in first panel. Bottom  right: $\chi^2$ values describing the goodness of fit of the best-fitting spectral age model. VLA BnA configuration C band (4-5 GHz) contours down to 4.0$\sigma$ (50.8, 87.9, 151.9, 262.6, and 453.8 $\mu$Jy per beam) are included in each image. The beam size of the radio contours is $1.6\times 0.5$ arcsec$^2$, and is shown in the bottom right corner (dashed black).}
    \label{fig:FullSrc_specagemap}
    \label{fig:FullSrc_specageerrmap}
    \label{fig:FullSrc_chisqmap}
\end{figure*}

 For the injection index, $\alpha_{inj}$, BRATS provides functionality to identify the best-fitting $\alpha_{inj}$ across the source. This is done by fitting the spectral aging model to each region with $\alpha_{inj}$ being an additional free parameter, and minimizing the sum of the model $\chi^2$ values over all  616 regions. For the VLA observations of \srcname, BRATS finds a best-fitting injection index of $\alpha_{\rm inj}=-0.70^{+0.03}_{-0.02}$ ($\chi^2=808.42$). With the best-fitting model we create a reconstructed model image of the source (Figure~\ref{fig:FullSrc_reconmap}; left), a residual map (Figure~\ref{fig:FullSrc_diffmap}; right), a spectral age map (Figure~\ref{fig:FullSrc_specagemap}; top), a spectral age error map (Figure~\ref{fig:FullSrc_specageerrmap}; middle), and a $\chi^2$ map corresponding to the goodness of fit of the best-fitting spectral aging model (Figure~\ref{fig:FullSrc_chisqmap}; bottom).

\subsection{Spectro-polarimetric properties of radio emission}
\label{sec:polar}
\begin{figure*}[ht!]
    \centering
    \includegraphics[width=0.49\linewidth]{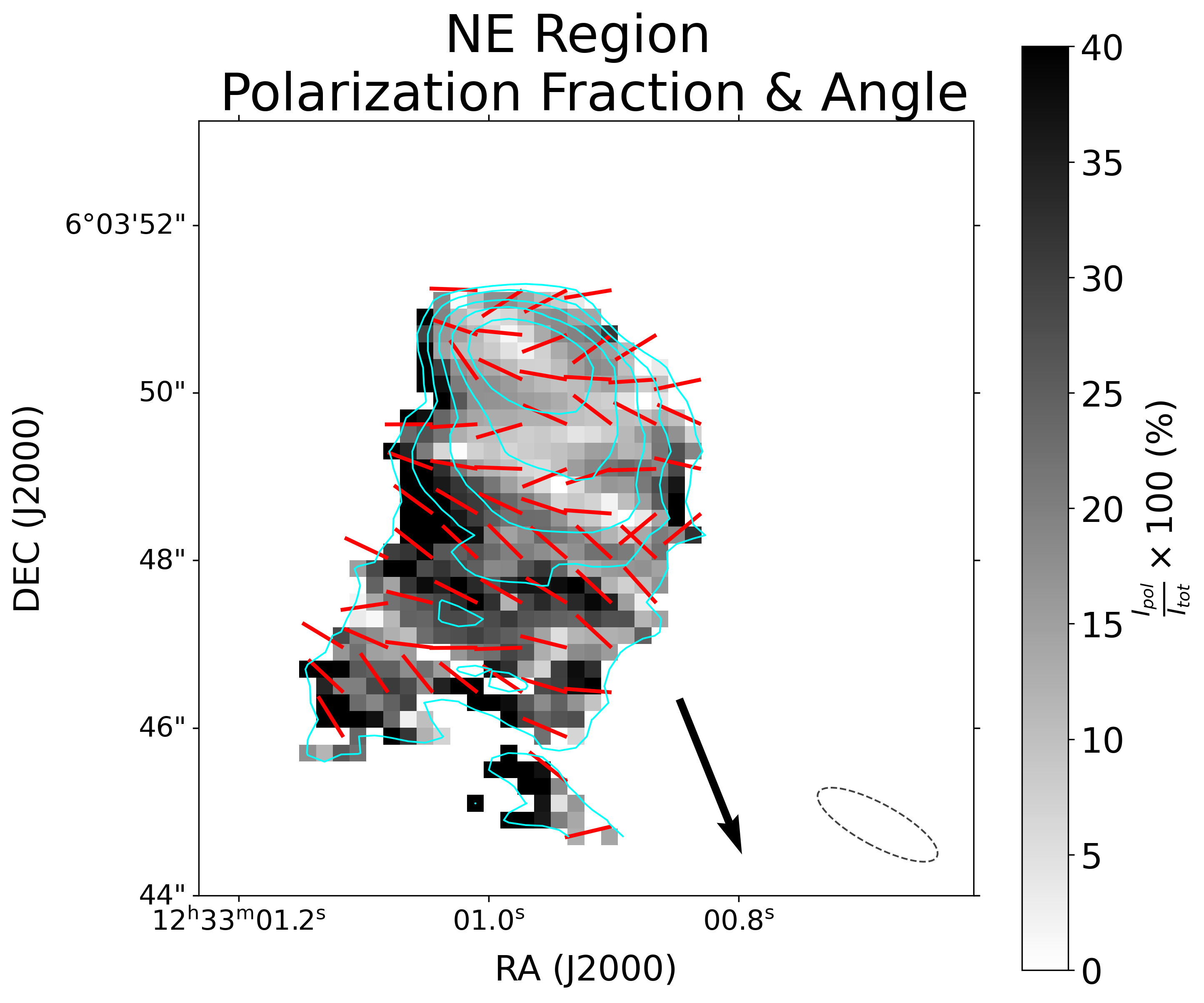}
    \includegraphics[width=0.48\linewidth]{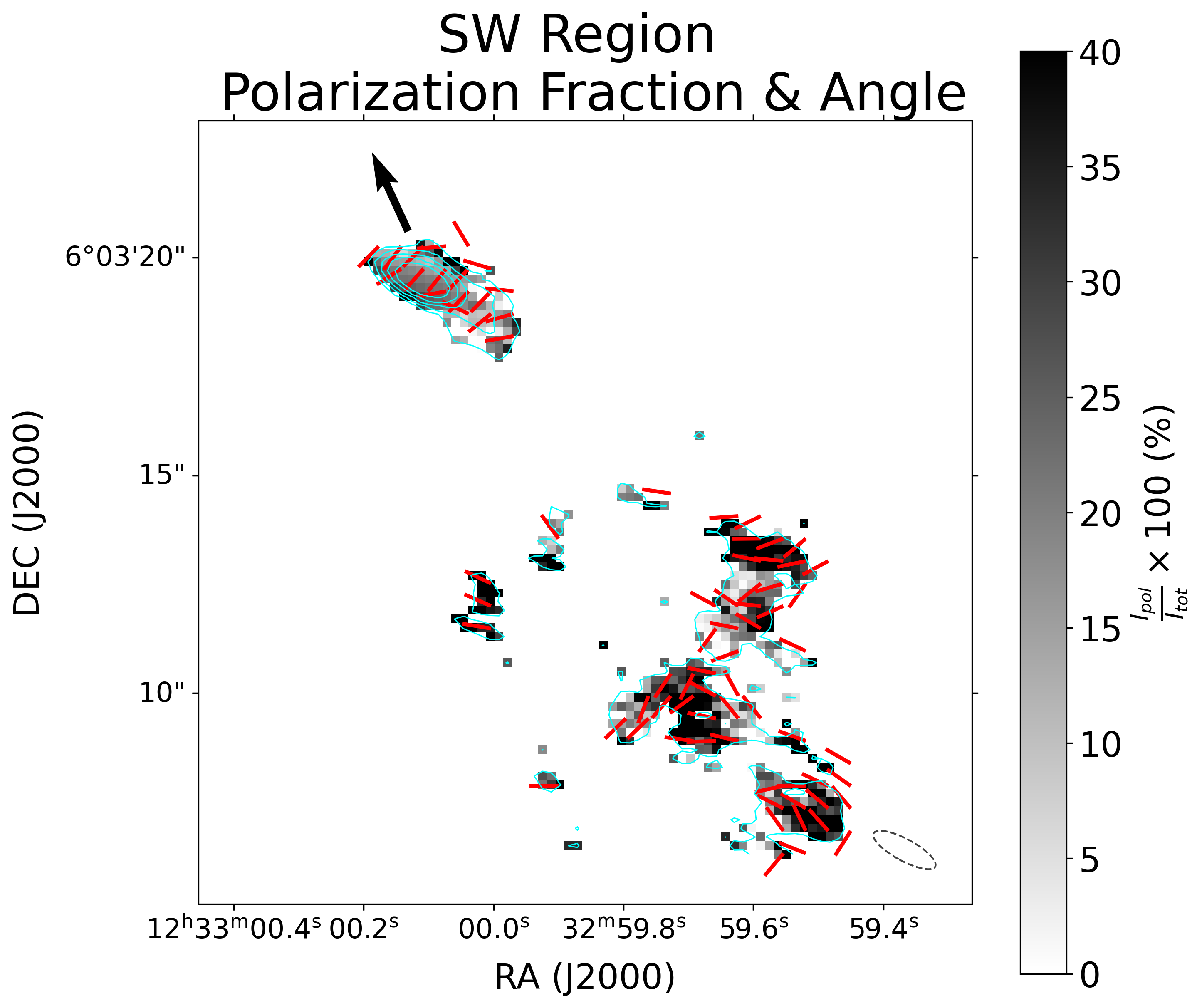}
    \caption{VLA BnA configuration C and X band (4-12GHz) polarization fraction map. The map of the angle of polarized intensity, rebinned such that there are $\sim$3 pixels/beam is shifted by 90 degrees, consistent with the definition of the projected magnetic field direction, and overlaid with red arrows. VLA BnA configuration broad band (4-12 GHz) contours down to 4.0$\sigma$ (24.3, 45.0, 81.9, 150.6, 276.8 $\mu$Jy/beam) are included in each image. The direction to the core from each hotspot is indicated by the black arrows. The beam size of the radio contours is $1.6\times 0.5$ arcsec$^2$, and is shown in the bottom right corner (dashed black).  Left: NE  region.  Right: SW  region. Note that the size of the images are different. }
    \label{fig:PolfracAng}
\end{figure*}

The resulting Stokes Q and U broadband (4-12 GHz) images were used to compute the polarized intensity, corrected for Ricean bias \citep[][]{Killeen1986}, and the resulting Stokes Q and U 1~GHz bandwidth images were used to compute the polarization angle and rotation measure of the observed emission. Specifically, the 1~GHz images were concatenated into a cube and the CASA Imagepol tool {\it po.rotationmeasure} was used to create a rotation measure (RM) map, corrected with a Galactic rotation measure of 6.6~rad/m$^2$ \citep{Hutschenreuter2022}, and a map of the polarized emission position angle (PA) at zero wavelength. A map of the broadband polarized intensity is shown in Figure~\ref{fig:PolfracAng} with the PA vectors, rotated by 90$^{\circ}$ to represent the $\vec{B}$ field rather than the actually measured $\vec{E}$ field, overlaid as red arrows. The average polarization fraction within the radio beam centered on the  mean of the 2D Gaussian fit to the locations of the NE  region and SW  region are 8.9\% with a standard deviation of 3.7\%, and 20.2\% with a standard deviation of 3.1\%, respectively. The Core is not detected in polarized emission in our VLA data.

From the 1 GHz bandwidth full Stokes cubes, we computed the fractional polarization as a function of frequency for each pixel in the polarized intensity image. A linear fit was performed in each pixel to determine the slope between the fraction of polarized emission and frequency (in GHz), and this slope was recorded for each pixel. The average slope value within the radio beam centered on the  mean of the 2D Gaussian fit to the locations of the NE  region is 0.002 with a standard deviation of 0.005, and of the SW  region is 0.008 with a standard deviation of 0.006, indicating no more than 1.5\% per GHz change in the fractional polarization with frequency for these radio components.  

With the RM map produced by CASA's {\it po.rotationmeasure} function, we investigate the change in the polarization position angle as a function of $\lambda^2$ to determine whether there are any Faraday screens present within the radio emitting source or along the line-of-sight to the observer. The average RM value within the radio beam centered on the  mean of the 2D Gaussian fit to the locations of the NE  region is -18.27~rad/m$^2$ with a standard deviation of 28.48~rad/m$^2$ ($\chi^2=0.72$), and of the SW  region is 17.98~rad/m$^2$ with a standard deviation of 12.84~rad/m$^2$ ($\chi^2=1.40$). The $\chi^2$ values for each component are the median $\chi^2$ values within the radio beam at that location. These RM values are consistent with negligible Faraday rotation in the hotspots of the source.

\subsection{ Host Optical Morphology}

\begin{figure}[t!]
\centering
    \includegraphics[width=3.5in]{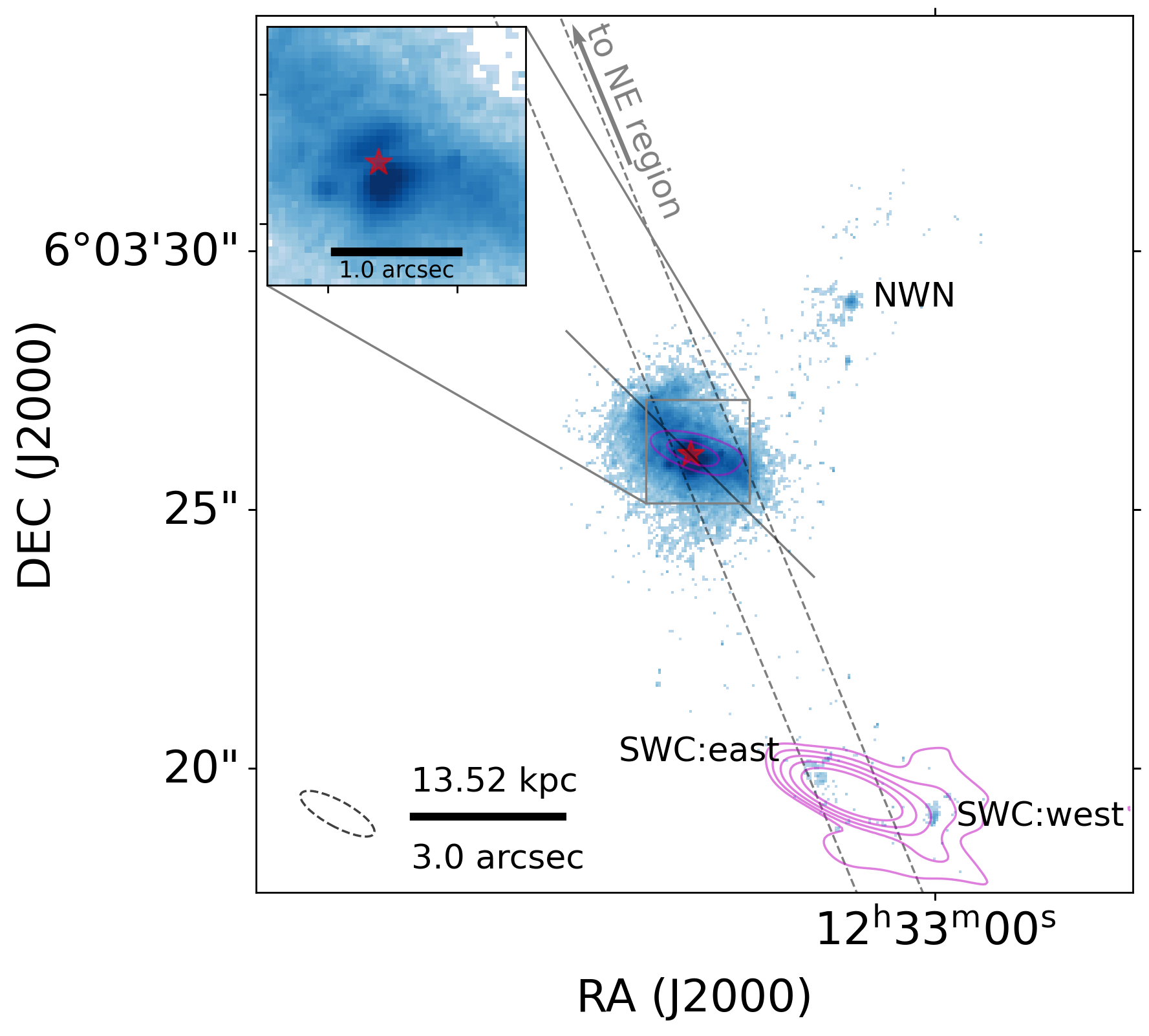}
    \caption{ A zoomed-in version of Figure \ref{fig:rdoCtr_optIm} focused on the host galaxy, detached EELN, and the source of brightest synchrotron emission in the SW region. The emission is manually masked to include only the  host galaxy nebula and the seemingly associated clouds, then these regions are masked down to 3$\times$ the off-source rms of the HST image. The color scales of the full optical image and the inset image are Log normalized between 0 and 0.5 and 0 and 1.0, respectively. The radio contour levels are drawn down to 4$\sigma$ (50.8, 87.9, 151.9, 262.6, and 453.8 $\mu$Jy/beam). The beam size of the radio contours is $1.6\times 0.5$ arcsec$^2$, and is shown in the bottom right corner (dashed black). The dashed rectangle represents the APO KOSMOS slit (1.18~\arcsec wide) aligned with the radio-jet axis, with the direction from the host galaxy to the NE region indicated by a black arrow. The red star indicates the image moment centroid.}
    \label{fig:OptIm_zoom}
\end{figure}

 Figure~\ref{fig:OptIm_zoom} shows a zoomed-in view of the HST F625W image, focused on the host galaxy, its EELR, and the peak of synchrotron emission in the SW region. The image reveals compact bright emission, co-spatial with the nucleus of the host galaxy, and asymmetrically distributed emission to the north and south of the nucleus. There is an elongated region of enhanced emission north of the nucleus spanning a direction that is nearly perpendicular to the radio-jet axis. Two smaller clumps of enhanced emission are visible on either side of the nucleus (see inset).

 The detached cloud, first identified in the APO spectra, is seen to the east of the SW region, and another detached cloud of similarly diffuse emission is present to the west of the SW region. We refer to these as South West Cloud:east (SWC:east) and South West Cloud:west (SWC:west), respectively. We also identify a compact source surrounded by diffuse emission to the north west of the host galaxy and we refer to this source as the North West Nebula (NWN). We manually mask the image to exclude all sources except the host galaxy, its EELR, and the three nearby regions of diffuse emission which may be associated with the EELR (SWC:east, SWC:west, and NWN), at a level of 3$\times$ the RMS in a 4~arcsec$^2$ source-free region. The centroid of the host galaxy is found with the raw image moments of the masked image, excluding the SWC:east, SWC:west, and NWN. The brightest parts of SWC:east, SWC:west, and the NWN lie at distances of 28.5~kpc, 37.8~kpc, and 19.46~kpc from the centroid, respectively. The APO KOSMOS longslit (Table~\ref{tab:observations}) aligned with the radio jet axis is included in Figure~\ref{fig:OptIm_zoom} for reference. 

\subsection{ Host Galaxy \& EELR Line Ratios}
We use the KOSMOS spectrum to calculate the [O~III]$\lambda$5007/H$\beta$ and [N~II]$\lambda$6584/H$\alpha$ ratios to determine the location of \srcname\ on the classic BPT \citep{Baldwin1981} diagnostic diagram.  The host galaxy of \srcname\ has BPT ratios of log$_{10}$([O~III]$\lambda$5007/H$\beta)=$0.98 $\pm$ 0.06 and log$_{10}$([N~II]$\lambda$6584/H$\alpha)=$-0.39 $\pm$ 0.08, placing  it above the theoretical maximum for star forming galaxies \citep{Kewley2001} and in the region dominated by AGN-ionized emission.  Although the [N II]$\lambda$6584 detection in the SWC:east was not significant enough for a flux measurement, we find an [O~III]$\lambda$5007/H$\beta$ ratio of log$_{10}$([O~III]$\lambda$5007/H$\beta)=$0.83 $\pm$ 0.12 for the SWC:east, placing it above the  theoretical maximum for star forming galaxies \citep{Kewley2001} if log$_{10}$([N~II]$\lambda$6584/H$\alpha$) $>$ -1.0. 
 Furthermore, \citet{Shirazi2012} showed that AGN can be more clearly separated from star-forming galaxies by using the He~II$\lambda$4686/H$\beta$ ratio in place of the [O~III]$\lambda$5007/H$\beta$ ratio since the high ionization potential of nebular He~II$\lambda$4686 (54.4~eV) makes it more sensitive to the presence of an AGN than [O~III]$\lambda$5007 (35.1~eV). For the host galaxy we find log$_{10}$(He~II$\lambda$4686/H$\beta$) = -0.28 +/- 0.09,  and for the SWC:east log$_{10}$(He~II$\lambda$4686/H$\beta$) = -0.33 +/- 0.35, placing them above the updated theoretical maximum for star forming galaxies \citep{Shirazi2012}, in a region where the AGN contribution to powering the emission is $>$75\%.

\section{Discussion}\label{sec:discussion}
 We have identified a new Green Bean galaxy hosting extended radio emission and a fractured EELR. We have combined optical long-slit spectra from APO and high resolution broadband radio observations (4-12~GHz) from the VLA with archival radio surveys (88~MHz-3.0~GHz) and pre-existing HST broadband imaging (WFC F625W) to investigate the nature of the EELR around \srcname, explore the feedback modes of the AGN, and disentangle the morphology of the large scale radio emission. In what follows, we discuss these results and suggest scenarios to explain the nature of \srcname.

\subsection{Powering Mechanism of the EELR}

 Our optical results suggest that the EELR around \srcname\ is likely powered by AGN photoionization. The SWC:east appears almost completely covered by the APO long-slit, suggesting the narrow line widths measured for this cloud (Table~\ref{tab:LineInfoDIS}) are characteristic across the SWC:east structure. Such narrow line widths combined with the high [O~III]/H$\beta$ and He~II/H$\beta$ ratios measured for both the SWC:east and the host galaxy are suggestive of AGN photoionization powering the emission, as opposed to shock heating (although this does not preclude the possibility of photoionization by a shock front that has not yet reached the illuminated region of the cloud, see Section~\ref{sec:Jet-Cloud}). In what follows, we assume that the EELR around \srcname\ is powered predominantly by AGN photoionization. Without spectroscopic information for the NWN, we cannot constrain its powering mechanism, but we note that the assumption of AGN photoionizations powering this source does not the affect our main results. 

\subsection{The Ionization Profile of the EELR}

 For gas that is heated via AGN photoionization, photoionization modeling \citep[e.g., CLOUDY][]{Ferland1998, Ferland2013, Ferland2017} can help map the rate of photons produced in recombination lines, such as H$\alpha$, to the rate of ionizing photons (Q$_{ion}$) intercepted by gas in the EELR. If the production of ionizing photons varies on timescales longer than the recombination time of the observed line, but shorter than the time it takes light to traverse the EELR, the record of this change will be encoded in the surface brightness distribution of the recombination line. Essentially, mapping the recombination line photons to AGN ionizing photons as a function of distance from the nucelus constrains changes in the rate of ionizing photons with time. While accurately performing this recombination line mapping requires information on the gas electron temperatures and densities \citep[see e.g., ][]{Finlez2022}, a lower limit to the expected rate of ionizing photons can be estimated by assuming that each recombination line photon corresponds to a single photoionization event (recombination balance).

 If the expected Q$_{ion}$ derived in this way is underestimated by an amount that is constant with distance from the nucleus, the Q$_{ion}$ vs. distance (lightyears) profile can still be interpreted in terms of how the ionizing photon rate has evolved over the light travel time of the EELR. The expectation that Q$_{ion}$ would be underestimated comparably at each distance assumes that the gas density of the EELR is relatively constant with distance from the nucleus. This assumption may be appropriate for EELRs which consist of photoionized tidal debris, particularly for the densest (brightest) clumps of gas. This so-called ionization history analysis has been demonstrated on HST narrow band observations of z$\sim$0.05 AGN hosting EELRs with photoionized tidal features \citep[e.g.][]{Keel2017}.

Specifically, \citet{Keel2017} mapped H$\alpha$ recombination emission to Q$_{ion}$ using narrowband imaging centered on the H$\alpha$ recombination line for a sample of fading AGN, and available narrowband imaging centered on [O III]$\lambda\lambda$4959,5007 (assuming [O III]/H$\alpha$=4) for a comparison sample of currently active AGN. They showed that the fading AGN sample has different behavior in the derived ionization history, compared to the active sample, with the former showing a distinct fall in expected ionizing flux from the outer EELR to the location of the AGN.

 Assuming photoionization equilibrium for a centrally located AGN surrounded by an EELR without unreasonable elongation along the line of sight, the ionizing photon rate Q$_{ion}$ can be estimated as: 

\begin{equation}
Q_{ion}  = \frac{N}{ fA}\frac{(4\pi~R^2)}{a^2}(4\pi~D_{L}^2)(1+z)   
\end{equation}

\noindent
 where N is the H$\alpha$ image count rate for each pixel in units of photons s$^{-1}$, A is the effective telescope area, R is the projected distance of each pixel from the nucleus (where it is assumed the production of ionizing photons occurs), a is the area of a single pixel as seen from the nucleus, D$_L$ is the luminosity distance for \srcname, z is the redshift, and f is the fraction of Hydrogen recombinations that lead to an H$\alpha$ photon. Following \citet{Keel2017}, we let $f=0.29$, which is appropriate for ``Case B" recombination.

 We manually mask the HST WFC F625W image to exclude all sources except the host galaxy, its EELR, and the three nearby sources, at a level of 2$\times$ the RMS in a 4 $arcsec^2$ source-free region. We then measure the image counts (electron s$^{-1}$) in each pixel of our masked image, corrected for the quantum efficiency of the HST WFC/F625W filter system ($\eta=$0.40 electrons photon$^{-1}$ at the observed wavelength of [O III]$\lambda$5007). The corrected counts (photon s$^{-1}$) are converted into H$\alpha$ image counts using the measured ([O III]$\lambda\lambda4959,5007$+H$\beta$)/H$\alpha$ ratios from the APO KOSMOS data, assuming other lines falling within the HST WFC F625W filter have negligible strengths compared to these three. For pixels lying within a projected distance of d$<$64,660 lightyears we find ([O~III]$\lambda\lambda4959,5007$+H$\beta$)/H$\alpha$ = 5.2 and for pixels lying above this distance we find ([O III]$\lambda\lambda4959,5007$+H$\beta$)/H$\alpha$ = 6.6, which we use to compute their ionizing luminosities.

\begin{figure}[t!]
\centering
    \includegraphics[width=3.5in]{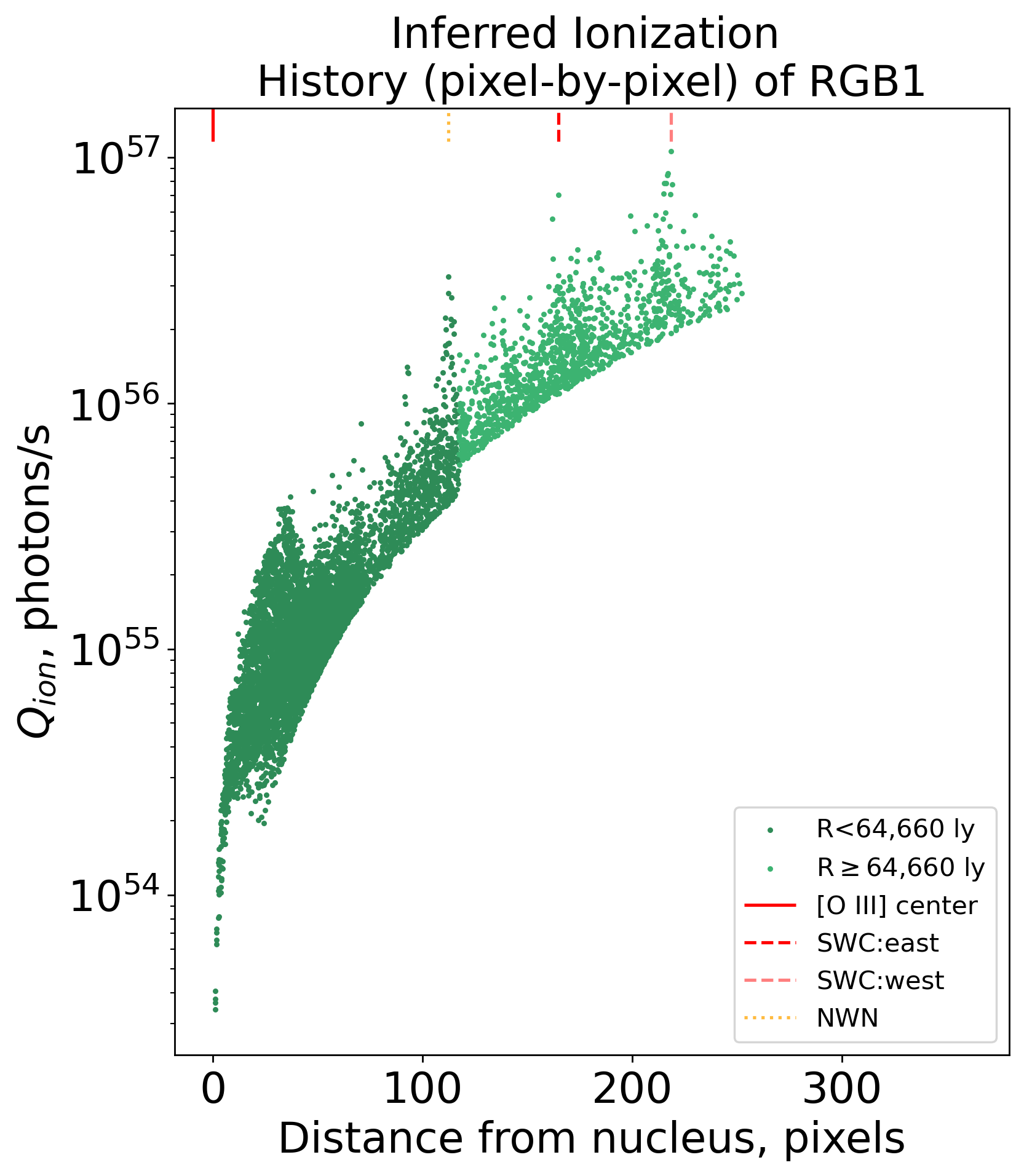}
    \caption{ Pixel-by-pixel inferred ionization history of \srcname. We mark the locations of the brightest pixel in the masked image (solid red line), and the brightest pixels in the SWC:east (dashed red), SWC:west (dashed pink), and NWN (dotted orange). The lower envelope is caused by the 3$\times$ off-source RMS cutoff used in the HST F625W image.}
    \label{fig:Qion_distance}
\end{figure}

The projected distance of each pixel (R) underestimates the true distances of each volume of gas (pixel) from the nucelus by a factor of $1/sin(i)$, where $i$ is the angle between the volume of gas and the line of sight. This means that the time delay between the arrival of nuclear emission to the observer and the arrival of reprocessed emission from a volume of the EELR lying at an angle of $i$ to the line-of-sight is: $(R/c)\times (1-cos(i))/sin(i)$ \citep{Keel2012b}, where c is the speed of light. The Q$_{ion}$ profile as a function of distance from the assumed location of the AGN (the brightest pixel) in the HST WFC/F625W image is shown in Figure~\ref{fig:Qion_distance}.

\begin{figure}[t!]
\centering
    \includegraphics[width=3.5in]{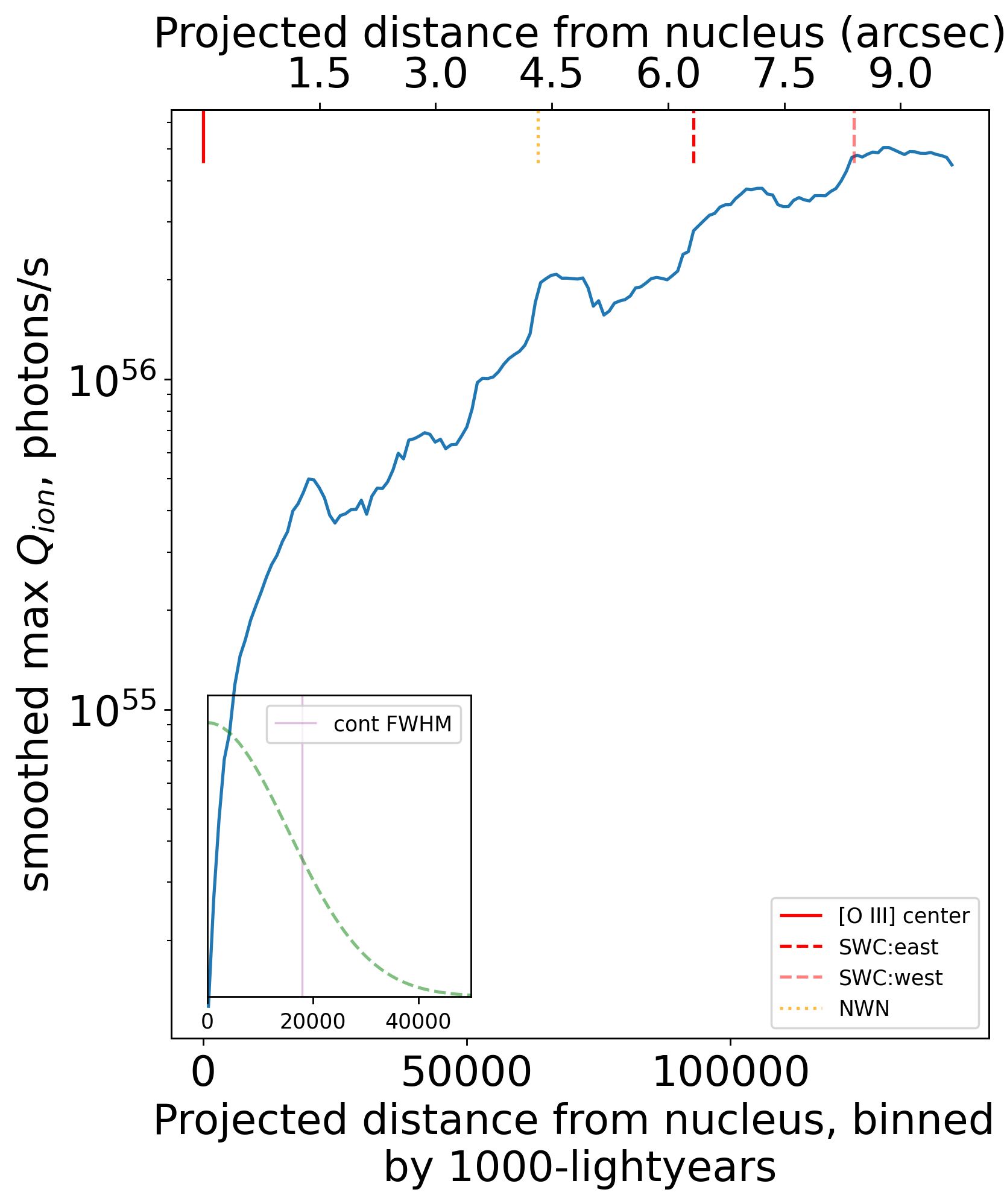}
    \caption{ Inferred Ionization History (Smoothed max Q$_{ion}$) of \srcname. We mark the locations of the brightest pixel in the masked image (solid red line), and the brightest pixels in the SWC:east (dashed red), SWC:west (dashed pink), and NWN (dotted orange). The dashed green Gaussian profile in the bottom left of the figure represents the size of the optical continuum, as suggested by the FWHM (purple) of the Gaussian profile fit to the spatial axis of the 2D spectrum. The Gaussian profile is arbitrarily normalized. }
    \label{fig:maxQion_smth}
\end{figure}

 The lower envelope of this ionization profile is caused by the 3$\times$ off-source RMS cutoff used for the analysis. The upper envelope corresponds to the densest (brightest) gas clumps at each distance, and thus gives us the strictest limits on the minimum Q$_{ion}$ required to power the H$\alpha$ recombination emission. To reconstruct this upper envelope, one can bin the pixels in distance by 1000 light years and select the max Q$_{ion}$ value in each bin. The temporal resolution of the ionization history derived in this way is limited by the recombination timescale of the observed line, which depends on the gas density in the EELR. This means the temporal resolution of the ionization history should degrade with distance from the source of ionizing photons \citep{Keel2017}. Following \citet{Keel2017}, we smooth our max Q$_{ion}$ values by 0.14$\times$R. The scaling of 0.14 is consistent with the recombination time scale ($\sim$8000 years) for the density (n$_{e}<$50~cm$^{-3}$) of the prototypical ionization echo, Hanny’s Voorwerp, which lies at a projected distance of $\sim$57,500 light years from its source of ionizing photons \citep{Lintott2009}. The smoothed ionization profile of \srcname\ (Figure~\ref{fig:maxQion_smth}) spans $\sim$2 orders of magnitude in Q$_{ion}$ and shows a clear decline from the outer to inner EELR, characteristic of the ``fading'' signature seen in \citet{Keel2017}.

 The change in Q$_{ion}$ with distance from the nucleus suggests that the ionizing photon rate of the AGN has declined by $\sim$2 dex over the past 0.15~Myr. Without comparable narrowband imaging centered on the H$\beta$+[O III]$\lambda\lambda$4959,5007 complex, we cannot constrain the continuum contribution to the broadband emission, therefore, we do not correct for the continuum, but note that removing it from the hot galaxy would bias the inner region towards lower Q$_{ion}$ estimates, further exaggerating the ``fading'' signature.

\subsection{Timescales of AGN activity}

The timescales associated with different states of accretion in AGN allow for detailed analysis of the contributions of radiative and kinetic feedback to the energy injected into the environment of the galaxy \citep{Fabian2012,Morganti2017}. Radiative mode feedback dominates during episodes of high-Eddington accretion, which produces radiation pressure driven winds and outflows. Kinetic mode feedback is thought to  dominate during the low-Eddington accretion phase via mechanical driven outflows,  which can persist as remnants in the system long after the AGN has transitioned into a different accretion state \citep{Schawinski2015}.   Evidence of these two modes of accretion has been observed simultaneously in some radiatively efficient AGN hosting jet-driven radio structures \citep[high excitation radio galaxies; HERGs,][]{Laing1994, Best2012}, and in some cases kinetic mode feedback has been observed to dominate in systems where one might expect radiative mode feedback \citep[i.e. in HERGs hosting jet-driven outflows,][]{villar-martin2017, Villar-Martin2021}.

\subsubsection{ The EELR as a Tracer of Radiatively Efficient Accretion}

In AGN, the total accretion lifetime is thought to be $\sim10-1000$~Myrs \citep{Soltan1982,Yu&Tremaine2002,Marconi2004,Hopkins2009}, but  may be broken up into smaller radiatively efficient accretion episodes of $\sim$0.1 Myrs \citep{Schawinski2015,King2015}. Based on our optical results we find that the minimum duration of the radiatively efficient accretion state  responsible for powering the production of ionizing photons in the \srcname\ system is of the order of  $\sim$0.1~Myr. Specifically,  the light travel distance between the nucleus of \srcname\ and the south-west clouds  suggests that ionizing photons released from the nucleus of \srcname\ would take at least  $\sim$0.09~Myr and $\sim$0.12~Myr to photoionize the extended gas  in SWC:east and SWC:west, respectively.  Furthermore, the ionization history analysis suggests that the rate of ionizing photons produced in the nucleus has declined by two orders of magnitude over the past $\sim$0.15 Myr.

 The emission used to derive these constraints traces AGN heating on scales comparable to the light travel distance of the EELR ($>$1000 light years), while a more direct view of the AGN as indicated by AGN-heated dust in the nuclear torus (10-100 lightyears) is traced by mid-IR (MIR) emission. Type 2 AGN with fading and rising rates of ionizing photon production can be identified by a comparison of their [O III] EELR luminosity vs. MIR luminosity. For this relationship, samples of active Type 2 AGN with radiatively efficient accretion flows will span a positively increasing locus of [O III] vs. MIR luminosity. AGN that fall above and below this track can be identified as potential fading and rising candidates, respectively \citep{Esparza-Arredondo2020, Pflugradt2022}. Using this technique, \citet{Schirmer2013} found that all 17 reported GB galaxies lie above this locus when compared to typical active Type 2 AGN and luminous obscured AGN, suggesting they host AGN with falling rates of ionizing photon production. 

 Using the APO KOSMOS spectrum, we find an [O III]$\lambda$5007 luminosity of log$_{10}$(L$^{KOSMOS}_{[O III]}$) = 42.2704  $\pm$  0.0042 for \srcname. Given that the APO spectra are not corrected for slit losses, we also estimate the [O III]$\lambda$5007 luminosity by using the ([O III]$\lambda\lambda4959,5007$+H$\beta$+continuum$_{5446\AA}^{7100\AA}$)/[O III]$\lambda$5007 ratio measured from the optical spectrum to estimate the contribution of [O III]$\lambda$5007 emission to the HST WFC/F625W image. In doing this, we find log$_{10}$(L$_{[O III]\lambda5007}$)= 42.7 $\pm$ 0.26. The estimates from the APO spectrum and HST image place \srcname's L$_{[O III]\lambda5007}$ $\sim$0.5~dex and $\sim$0.1~dex below the least [O III]$\lambda$5007-luminous GB galaxy, respectively. In the MIR, the Wide-field Infrared Survey Explorer \citep[WISE,][]{Wright2010} 22$\mu$m luminosity for \srcname\ is log$_{10}$(L$_{22\mu m}$) = 44.28 $\pm$ 0.16, consistent with the GB galaxies. This suggests that \srcname\ would not be selected as a fading AGN candidate by this type of [O III] vs. MIR inspection.

 One reason why the fading behavior of \srcname\ is captured by the ionization profile analysis but not the [O III] vs MIR comparison could be that \srcname\ does not have as much extragalactic gas available to be illuminated by the AGN, compared to the GB galaxies. If this is the case, this suggests that using an inspection of [O III]$\lambda$5007 vs. MIR emission to identify fading candidates may bias the selection towards AGN with more massive EELRs outside their host galaxy.

 The lack of fading behavior in the [O III]$\lambda$5007 vs. MIR inspection may also be due to differences in the host galaxy properties of \srcname\ and the GB galaxies. The GB galaxies have MIR emission that follows a power law, suggesting the emission is nuclear in origin \citep[i.e. AGN heated,][]{Schirmer2013}. In \srcname\ the MIR WISE flux densities (F$_{\lambda}$) show a decrease in brightness between the rest frame wavelengths 2.6-3.5 $\mu$m before monotonically rising between 3.52-16.9 $\mu$m, suggesting that the AGN is not dominating the MIR emission. Given that the 22$\mu$m luminosity of \srcname\ is comparable to the GB galaxies, this may indicate higher rates of star formation or a particularly dusty interstellar medium (ISM) in \srcname. In this case, a decline in the MIR emission associated with AGN-heated dust may not be sufficient to allow \srcname\ to stand out in an [O III]$\lambda$5007 vs. MIR inspection. However, the dominance of the AGN in powering the EELR emission via photoionization combined with the low AGN contribution in the nucleus as inferred from the MIR emission further supports a decline in the Q$_{ion}$ of the AGN over the past $\sim$0.15~Myr.

\subsubsection{The Spectral Age of the Radio Emission}

 Interpretation of the radio spectral aging analysis is not as clear. The spectral index of the radio core is steep ($\alpha^{10.5}_{4.5}=-1.02\pm0.09$), while typical spectral indices for active radio galaxy cores are much flatter, generally $\alpha_{GHz}>-0.5$ \citep{Burns1982, Hardcastle2008}. This observed steepness could be caused by a recent change in the accretion rate of the AGN in \srcname.

 The spectral slopes of the compact components in the NE and SW regions are also steep, but consistent with the typical spectral indices measured in active hotspots and fresh lobes of radio galaxies where the median is typically $-0.75$ to $-0.85$ \citep[][]{Herbig1992,Kharb2008,ODea2009,Vaddi2019}, and may be as steep as $-1$ in the case of a strong termination of the jet (``high-loss" hotspots) where the electrons accelerated by the jet termination lose most of their energy before being transported into the lobes \citep{Meisenheimer1989}. However, if such strong shock terminated in the SWC complex, then this seems at odds with the narrow line widths in the EELR east of the presumed SW region hotspot (SWC:east; but see Section~\ref{sec:Jet-Cloud} for discussion of a potential jet-cloud interaction). Alternatively, if the steep spectrum of the radio core is due to a recent change in the accretion rate, the somewhat steeper spectral slopes of the NE and SW compact components as compared to the typical values of active radio lobes may be caused by the initial decline in jet power.

The spectral modeling revealed low age regions in the center of the NE BRATS region and along the southern edge of the SW BRATS region (Figure~\ref{fig:FullSrc_specagemap}, top). These low aged plasma regions, which may indicate where the true hot spots most likely are, span ages of 0-3~Myrs. Visual inspection of the model fits in these low aged areas suggests that the regions where the ages are exactly 0~Myr are driven by noisy data and poor model fits (note the higher spectral age errors and chi squared values associated with the zero-aged regions). Furthermore, our visual inspection also determined that the low age plasma regions with ages above 0~Myrs are associated with adequate model fits, but high spectral age errors ($2-3$~Myrs). This makes it it impossible to determine whether the hotspots of \srcname\ are still active, or remnant with an off time of $\sim2-3$~Myrs.

The average age for all pixels included in the NE  BRATS regions are  6.20~Myrs with standard deviation  2.54~Myrs, and in the SW  BRATS regions  6.37~Myrs with standard deviation  1.55~Myrs.  These averages suggest that the duration of the phase responsible for powering the radio jet in \srcname\ is at least $\sim$6~Myr, but likely much longer given that these averages are more heavily weighted towards hotspot emission (lowest ages) rather than emission from the diffuse lobe plasma (oldest ages). Resolved spectral age modeling across all scales of the low frequency emission is imperative to estimating the true age of the oldest plasma. 

\subsubsection{Timelines of the AGN feedback modes}

Our ionization history analysis for \srcname\ suggest that this AGN was previously active in a high-accretion state, i.e. the efficient accretion state associated with high Eddington-scaled accretion rates and a high production rate of ionizing photons. Then,  over the course of 0.15~Myr, the  high-Eddington accretion rate declined, leaving behind an EELR ionization echo.  Being further from the nucleus, the large scale jet-induced radio emission associated with \srcname\ probes longer timescales of the AGN activity.  The low age regions found by the spectral aging analysis are consistent with active jet termination in the presumed hotspots, while the spectrally steep, faint emission from radio core suggests the jet-production potentially being in a phase of switching off.

Combined multiwavelength studies are a powerful tool in deciphering accretion stages and timescales in AGN \citep[e.g.][]{Fabbiano2022,Alcorn2023}. Recently X-ray, UV, optical, IR, and radio observations have been combined to characterize the accretion history of the AGN in IC~2497, which hosts the prototypical ionization echo, ``Hanny's Voorwerp" \citep{Lintott2009},  hence being particularly relevant here. Multi-wavelength studies of this system have revealed that the AGN in IC 2497 shows evidence of  both a remnant \citep[t$_{spec}>$100~Myr;][]{Smith2022} and a restarted \citep{Jozsa2009,Rampadarath2010} phase of jet production, along with signs of kinetic feedback  currently acting on the host galaxy \citep{Keel2012b, Sartori2016}. The highly ionized gas in ``Hanny's Voorwerp" is likely associated with an episode of radiatively efficient AGN activity that ended $\sim$0.1~Myrs ago, sometime between the two kinetic mode episodes. The general picture revealed by combining these studies is one where the AGN in IC~2497 has alternated between radiatively inefficient and efficient accretion states (possibly dipping into quiescent states) for the past $\geq100$~Myrs.

 If the jet power responsible for the radio emission in \srcname\ is declining then the picture painted by combining our results for \srcname\ suggests that it may have been a HERG whose jet power and ionizing photon rate have declined over the past $\sim$2--3~Myr and $\sim$0.15~Myr, respectively. In this case, it is not clear whether the AGN transitioned from an accretion state associated with radio jet production to an accretion state associated with ionizing photon production before declining in nuclear energy production all together, or whether the rise and fall in nuclear ionizing photon rate occurred alongside the radio jet production phase (i.e. feedback modes acting simultaneously). On the other hand, if the radio emission in \srcname\ is associated with active jet production, then our results suggest that \srcname\ was a HERG that transitioned to a low excitation radio galaxy (LERG) in the past $\sim$0.15~Myr.

\subsection{A Potential Jet-Cloud Interaction} 
\label{sec:Jet-Cloud}
 We observe a superposition of the VLA-detected SW region and the SWC complex (Figure~\ref{fig:OptIm_zoom}) that is suggestive of a jet-cloud interaction. Hydrodynamical simulations indicate that interaction of both non-relativistic and relativistic jets with dense clouds in extragalactic radio sources may cause significant changes in these components including cloud destruction or displacement, and jet bending or even termination especially in the case of weak jets \citep{Higgins1999,Wang2000,Choi2007}.

In principle, the observed superposition of the VLA-detected SW  region and the  SWC complex may be incidental. The lack of broad line widths in the  SWC:east spectrum can support such a scenario. However, one explanation for the narrow line widths observed in the cloud could be due to the slit of the optical spectrum not being fully aligned with the SW  region hotspot in our observations (Figure~\ref{fig:OptIm_zoom}). In 3C~277.3 \citep[Coma A;][]{Miley1981}, which is undergoing jet-cloud interactions on both sides of its nucleus \citep{vanBreugel1985}, integral field spectroscopic  (IFS) observations of the system found lower H$\alpha$ line widths for gas slightly offset from the northern jet hotspot \citep{Inarrea2003}. The authors suggest that this gas has not yet been disturbed by the radio jet, but is being ionized by radiation produced in the shock \citep{Inarrea2003,Worrall2016}.  This suggests that we should see evidence of an on-going shock in the SW region hotspot.

Based on hydrodynamical simulations jet--cloud interaction would produce sharply increased synchrotron emission at the location \citep{Choi2007}. The coincident radio components in the VLA observations suggest such an incidence. However, in our VLBA observations of the SW  region we do not detect any emission from the hotspot or jet knots, with a 3$\sigma$ upper limit of 69.3~$\mu$Jy/beam. While actively replenished hotspots should be of a brightness temperature detectable at VLBI angular resolutions,  it is possible for them to be resolved out. Based on the shortest baseline for the VLBA observations, a non-detection of an active hotspot could imply that its size is larger than 275~mas (1.24~kpc at the redshift of \srcname). However, htspot sizes detected in VLBI experiments are observed to be 50~pc -- 2.5~kpc in size \citep[e.g.][and references therein]{Tingay2008}. Alternatively, the hotspot may simply be too faint to be detected in our observations with their brightness temperature limit of $7.1\times10^4$~K. In principle this is enough to detect hotspots of powerful radio galaxies at mas scales \citep[e.g.][]{Lonsdale1998,Young2005,Tingay2008}, but our radio galaxy is significantly fainter and less powerful.

An alternative suggestion , in the case where the jet power has declined over the past $\sim$2--3~Myr, is the possibility that shocks induced in the cloud $\sim$ 2--3~Myrs ago due to the termination of the SW jet may have dissipated quickly enough to result in such quiescent cloud kinematics seen today. Such a scenario is also suggested as a possibility for the IC~2497/``Hanny's Voorwerp" system by \citet{Smith2022} in order to explain the quiescent kinematics of ``Hanny's Voorwerp" given that the large-scale radio jet is believed to have punched a ``hole" through the nebula  $>$100~Myr ago. Deeper, higher resolution IFS observations may reveal whether the cloud is being shocked by the jet by estimating whether the resolved electron temperatures and densities across the cloud are consistent with shock models \citep[][]{Davies2015}.

\subsection{Radio Morphology of \srcname}
\label{sec:morph}
\srcname\ was first spotted due to its hybrid-like radio morphology appearing in mid-resolution radio images, hence a brief discussion on this aspect is called for here. HyMoRS are an interesting class of object providing insight in the Fanaroff-Riley dichotomy debate \citep[nature vs. nurture, e.g. ][among many others]{Reynolds1996, Meier1997, Laing2002, Laing2014, Meliani2008, Saripalli2012}. With its HyMoRS-like, asymmetric apparent radio structure, \srcname\ could in principle be considered a member of this morphological class. Radio morphology of the NE  region ($\sim26\arcsec$ distance from the radio core) does resemble a classical FR II source, with fairly compact structure located at the outermost part of the lobe, presumed to be a jet termination point, and a more diffuse lobe emission trailing back towards the core. The most compact component of the SW  region peaks in brightness at a distance $\sim 7.3\arcsec$ from the radio core, with more diffuse emission extending to the southwest, away from the host galaxy and the radio core. The polarization properties of the hot spots display signatures typical for FR~II types where the polarization angle is near perpendicular to the jet direction (signature of plasma compression, Figure~\ref{fig:PolfracAng}). 
Based on these results and the asymmetry of the radio source, \srcname\ could simply be an FR~II radio galaxy viewed at an angle to our line of sight. 

A suggestion has been recently put forward that some radio galaxies with the apparent HyMoRS type and clear hot spots in each of the lobes may be intrinsically FR~II sources with bent jets seen in projection \citep{Harwood2020}. However, both radio lobes of \srcname\ have similar, steep average spectral indices ($\alpha^{12\rm GHz}_{4{\rm GHz}}\simeq-0.9$; Figure~\ref{fig:radioSEDs}), and we do not see any significant depolarization in either  region (Section~\ref{sec:polar}); both of these effects  would be expected to be present at least in some degree if the bent-jet FR~II model applied to \srcname\ \citep[specifically the one in which jets are oriented away from the observer, cf. FR~I coun\-ter-jet viewing angle, Figure~4 in][]{Harwood2020}.  However, absence of these effects does not necessarily rule out the possibility of projection.


It has also been previously discussed that HyMoRS may potentially be FR~II type radio galaxies seen in a transient stage, i.e. during the radio source switching off, restarting, or undergoing episode of increased activity, in combination with projection effects  \citep{Gopal-Krishna1996, Marecki2012, Kapinska2017, Harwood2020}. For a radio source viewed at an angle to the line of sight, the central engine output modulation and differential light travel time between the far and near hotspots (relative to the observer) may temporarily produce an apparent HyMoRS.  For example, for a radio source seen in projection  that has undergone augmentation of activity in producing radio jets the near radio lobe and  its hotspot should be brighter than the far one. This is what we observe in \srcname\  where the NE hotspot is nearly twice as bright as the SW one, suggesting such an event could have happened recently. However, the asymmetry in the brightness of hotspots may be also due to other causes such as difference in the gas density around the source, with the brighter hotspot produced in higher density environment (but see the discussion in previous section on the possible gas--jet interaction in the SW region). 

Polarization properties of the SW region radio emission indicate a highly turbulent medium (Figure~\ref{fig:PolfracAng}); whether the lobe is seen in projection or it is truly trailing behind the fading hot spot away from the radio core, the environment on the southern side of the \srcname\ seems to be inhomogeneous, possibly denser, and more turbulent as compared to the northern side. While our results presented here are inconclusive regarding the viewing angle of the radio emission in \srcname, should the radio galaxy prove to not be projected, then the HyMoRS structure would still be caused by an early FR type II jet termination rather than the jet being intrinsically an FR~I type on one side.

\section{Conclusions}
We have presented a multi-wavelength investigation into the  nature of the \srcname\ system, including the powering mechanisms of its EELR, and the feedback modes of its AGN. We summarize our results as follows:

\begin{enumerate}
     \item \srcname\ is a new Green Bean galaxy hosting extended radio emission and an EELR that is most likely photoionized by an AGN. 
    \item The minimum duration of the radiatively efficient accretion state in \srcname\ is given by light-travel estimates to be  $\sim$0.1~Myr, while the ionization profile analysis indicates a decline in the rate of ionizing photons produced in this state over the past $\sim$0.15~Myr.  The spectral age estimates of the extended radio emission suggest the  duration of the jet production phase of the AGN  is at least $\sim$6~Myr, but are inconclusive regarding whether the jet production phase is still ongoing. Combined, these results suggest that \srcname\ was in a phase of jet production prior to the radiatively efficient accretion phase traced by the current EELR emission. It is unclear whether \srcname\ was a HERG that has transitioned into either a LERG or an inactive galaxy over the past $\sim$0.15~Myr, or whether the extended radio and optical emission trace distinct accretion phases that occurred in succession. 
    \item The SWC:east along the radio axis is aligned with the hotspot in the SW  region and displays a blue shift relative to the host galaxy. The cloud does not display any other evidence of turbulence (e.g. large line dispersion). Additional optical IFU and X-ray observations can shed light on whether the jet powering the SW  region terminated in the complex, or whether we are observing a chance alignment between the two. 
    \item While the VLA continuum total intensity image of the source suggest a hybrid radio morphology, the spectro-polarimetric radio observations suggests that both hotspots are produced as a result of the jet terminating in the external medium, consistent with both components being of FR~II morphology. Whether the HyMoRS appearance of \srcname\ is caused by an intrinsic FR type II source seen in projection or by an early termination of the southern jet into a more dense environment is unclear. 
\end{enumerate}

 \srcname\ is an interesting system that warrants multiwavelength follow-up on resolved scales to better understand the history and future of the host galaxy and the fading AGN within.

\section*{Acknowledgements}
The authors would like to thank Preshanth Jagannathan, Rick Perley, Natalie Wells, and Daniel Gondines for helpful discussions.  

The initial observation of \srcname\ was undertaken as part of the National Radio Astronomy Observatory (NRAO) VLA Summer Program funded by the National Science Foundation (NSF) using director's discretionary time. The authors would like to acknowledge Jacob Hetrick for their help with the initial data calibration. Support for the analysis of these observations was provided by the NSF through the Grote Reber Fellowship Program administered by Associated Universities, Inc./National Radio Astronomy Observatory. The National Radio Astronomy Observatory is a facility of the National Science Foundation operated under cooperative agreement by Associated Universities, Inc. 

\bibliography{sample63}{}
\bibliographystyle{aasjournal}



\end{document}